\documentclass[final,3p,times,twocolumn]{elsarticle}

\usepackage{graphicx}
\usepackage{amssymb}

\usepackage{amsmath}

\usepackage{array}
\usepackage{longtable}

\usepackage{lineno}

\journal{Nuclear Data Sheets}

\begin{document}

\begin{frontmatter}


\title{Remeasurement of the $^{239}$Pu(n,f)/$^{235}$U(n,f) Cross-Section Ratio with the NIFFTE fission Time Projection Chamber Using Vapor-deposited Targets}



\cortext[corresponding]{Corresponding author}

\author[LLNL]{D.H.~Dongwi}\corref{corresponding}
\ead{dongwi.dongwi@stonybrook.edu}

\author[LLNL]{L.~Snyder}\corref{corresponding}
\ead{snyder35@llnl.gov}

\author[CAL]{V.~Aguilar}
\author[CAL]{N.~Androski}
\author[LLNL]{M.~Anastasiou}
\author[LLNL]{N.S.~Bowden}
\author[OSU]{A.~Chemey}
\author[LLNL]{T.~Classen}
\author[CAL]{J.E.~Fuzaro Alencar}
\author[CSM]{U.~Greife}
\author[CSM]{M.~Haseman}
\author[ACU]{L.D.~Isenhower}
\author[CAL]{J.L.~Klay}
\author[OSU]{W.~Loveland}
\author[LLNL]{M.P.~Mendenhall}
\author[LLNL]{M.~Monterial}
\author[OSU]{M.~Silveira\fnref{ORNL}}
\author[LANL]{C.~Prokop}
\author[ACU]{T.S.~Watson}
\author[OSU]{L.~Yao}

\author[]{\protect\\(The NIFFTE Collaboration)}

\address[LLNL]{Lawrence Livermore National Laboratory, Livermore, CA 94550, United States}
\address[LANL]{Los Alamos National Laboratory, Los Alamos, NM 87545, United States}
\address[ACU]{Abilene Christian University, Abilene, TX 79699, United States}
\address[CAL]{California Polytechnic State University, San Luis Obispo, CA 93407, United States}
\address[CSM]{Colorado School of Mines, Golden, CO 80401, United States}
\address[OSU]{Oregon State University, Corvallis, OR 97331, United States}
\address[UC]{University of California, Davis, CA 95616, United States}

\fntext[ORNL]{Current Address: Oak Ridge National Laboratory, Oak Ridge, TN 37830, United States}

\date{\today}
	
\begin{abstract}
 The NIFFTE fission Time Projection Chamber (fissionTPC) has been used to measure the $^{239}$Pu(n,f)/$^{235}$U(n,f) cross-section ratio for neutron-induced fission in the range of 0.1 -- 100 MeV, with high precision. A white neutron source was provided by the Los Alamos Neutron Science Center, where the experiment was conducted as a remeasurement to evaluate a roughly 2\% discrepancy of the previous fissionTPC results with ENDF/B-VIII.1. A detailed accounting of measurement uncertainties was performed, based on the fissionTPC's novel ability to provide three-dimensional reconstruction of fission-fragment ionization profiles. 
 Current results obtained using a vapor-deposited, highly uniform $^{239}$Pu target, in comparison to the measurement published in 2021, where a $^{239}$Pu electroplated target was used, are presented and discussed.
 The remeasurement presented here is in agreement with the previous fissionTPC result within measurement uncertainties.
\end{abstract}

\end{frontmatter}

\tableofcontents{}


\newcommand{\ie}{\emph{i.e.}}
\newcommand{\eg}{\emph{e.g.}}
\newcommand{\etc}{\emph{etc.}}
\newcommand{\daq}{DAQ}
\newcommand{\pu}[1][239]{$\mathrm{^{#1}}$Pu}
\renewcommand{\u}[1][235]{$\mathrm{^{#1}}$U}
\newcommand{\hyd}[1][1]{$\mathrm{^{#1}}$H}  
\newcommand{\talpha}{$\alpha$}
\newcommand{\talphas}{$\alpha$s}
\newcommand{\ftpc}{fissionTPC}
\newcommand{\mev}{MeV}
\newcommand{\mus}{~$\mu$s}
\newcommand{\massunits}{\ensuremath{\mathrm{\mu g /cm^{2}}}}

\section{Introduction}
\label{sec:introduction}

Neutron-induced fission reactions have significant influence on a variety of fields of study, ranging from stellar nucleosynthesis to nuclear reactor design and safety. These latter systems have become increasingly dependent on models and simulations which require neutron-reaction cross-sections as inputs. Advances in computing power and theory have made these models more detailed and precise. As a result, there has been reinvigorated interest to better understand and quantify nuclear data uncertainties that feed into calculations and inform theoretical constructs. 
Modern data evaluation techniques require significant detail regarding uncertainty quantification \cite{Smith2012}. Various experimental datasets, considered in former nuclear data evaluations, have demonstrated significant underestimation of their reported uncertainties, leading to discrepancies between the observed values and the evaluated ones that are larger than the reported uncertainties. The cause of such discrepancies may be attributed to the so-called ``Unrecognized Sources of Uncertainties'' (USU), as introduced by Capote \emph{et al.}~\cite{Capote2020}. Recent efforts by the nuclear data evaluation community have expressed the need to better understand, quantify and report in detail all sources of uncertainty. At the same time, there exists a need for precision neutron-induced fission cross-section measurements, at the sub-percent level, on fissile nuclei such as $^{239}$Pu (see also recent work by Ref.~\cite{Mumpower:2023bhz}).

\begin{figure}[ht]
 \centering
 \includegraphics[width=1.\columnwidth]{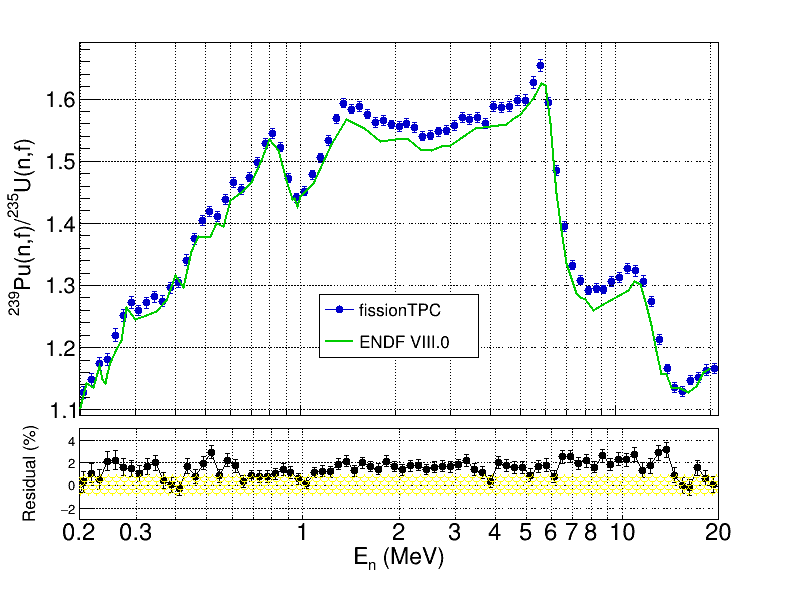}
 \caption{The  \ftpc{} measurement of the \pu(n,f)/\u(n,f) cross-section ratio as a function of neutron energy compared to the ENDF/B-VIII.1 evaluation published in 2021~\cite{Snyder2021jor}. The residual ((data-ENDF)/data) between the data and evaluation is shown in the lower panel, where the yellow band is $\pm$ 1\% from a zero residual. The error bars on the residual are from the \ftpc{} data only.}
 \label{fig:fy17TPC_ENDF8}
\end{figure}

The NIFFTE collaboration constructed a fission time projection chamber (fissionTPC) to measure neutron-induced fission cross-section ratios as a function of incident neutron energy, with total uncertainties less than 1\% (both statistical and systematic). By comparison, conventional measurements using ionization chambers are limited to 3–5\% total uncertainty \cite{Staples1998, Tovesson2015}.
A fissionTPC measurement of the $^{239}$Pu$(n,f)/^{235}$U$(n,f)$ cross-section ratio was completed on data collection in 2017 and detailed in two publications~\cite{Snyder2021jor} and~\cite{Monterial2021} in 2021, with the results uploaded to the EXFOR and Neutron Data Standards Database databases~\cite{neudecker2021}. The results from measurement~\cite{Snyder2021jor} differed from the ENDF/B VIII.1 evaluation by just under 2\% over all neutron energies, as shown in Fig.~\ref{fig:fy17TPC_ENDF8} from Ref.~\cite{Snyder2021jor}.

While slight disagreements between the fissionTPC results and the ENDF evaluation as well as other experiments were expected, the size of the discrepancy and how it remains systematically consistent as a function of incident neutron energy was indicative of a systematic offset in the measurement. Extensive analysis and validation studies carried out in Ref.~\cite{Snyder2021jor} led to the consideration of two possible causes, both related to the target. The first potential cause was the normalization measurement, which was conducted only after the cross-section ratio neutron-beam data were collected. The normalization, or target mass ratio, was obtained with a combination of two auxiliary measurements: a mass spectrometry measurement to determine the isotopic content of the targets, and a low geometry alpha counting experiment~\cite{Monterial2021}.
Damage from the beam during data-taking, and perhaps mishandling of the target, could have affected the final normalization value.  There was no direct evidence of this cause beyond the systematic nature of the discrepancy with ENDF.

\begin{figure}[ht]
 \centering
 \includegraphics[width=1.\linewidth]{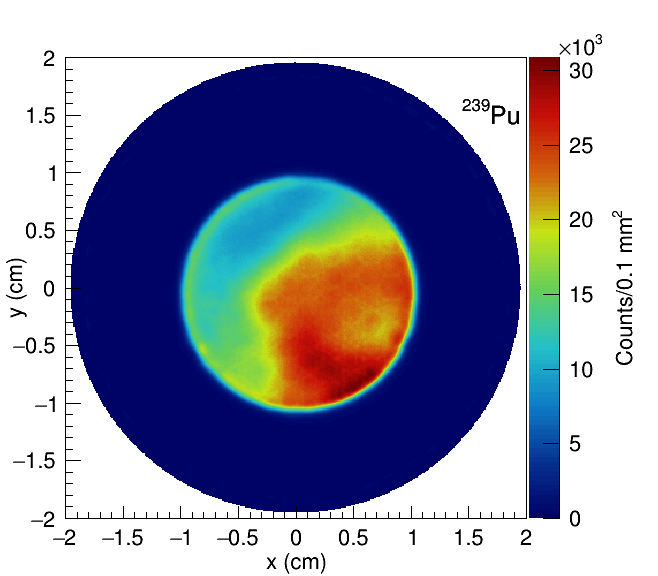}

 \includegraphics[width=1.\linewidth]{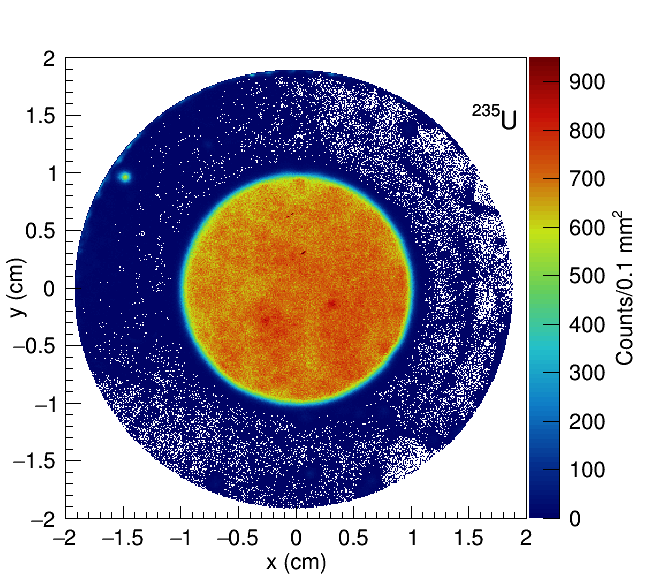}
 \caption{Image of \talpha-particle track start vertices for plutonium (top) and uranium (bottom) target deposits from the 2021 published data~\cite{Snyder2021jor}.}
 \label{fig:vertDist}
\end{figure}

The second possibility comes from the nonuniformity of the electroplated  $^{239}$Pu target in the first measurement, which is shown in Fig.~\ref{fig:vertDist}. Target nonuniformity coupled with a spatial nonuniformity in the beam flux necessitated a correction on the order of 3–5\%. 
While this is a large nonuniformity, the principle aim of the fissionTPC is to quantify potential systematic errors of this type and it was well within the instrument's capability to do so. 
Several validations detailed in Ref.~\cite{Snyder2021jor} outline that this correction was handled properly.
Furthermore, the correction involving beam-target nonuniform spatial distribution varied as a function of incident neutron energy, whereas the aforementioned discrepancy with ENDF appears to be largely systematic across all energies. 

For these reasons, a new measurement with a uniform, vapor-deposited $^{239}$Pu$/^{235}$U target~\cite{Loveland2016} was undertaken to directly address the $\sim$2\% discrepancy.

\section{Experiment}
\label{sec:experiment_description}
The $^{239}$Pu$(n,f)/^{235}$U$(n,f)$ cross-section ratio experiment was conducted at the Los Alamos Neutron Science Center (LANSCE) Weapons Neutron Science (WNR) facility on the 90L flight path \cite{LANSCE}. The LANSCE facility makes use of an 800 MeV proton accelerator to provide a pulsed white neutron source. For this experiment LANSCE delivered 100 Hz pulses, each approximately 625 $\mu$s in length, containing micro-pulses separated by 1.8 $\mu$s. Incident neutron energies were calculated by taking advantage of a neutron time-of-flight measurement as described in Ref.~\cite{Snyder2021jor}.

The \ftpc{} is a two-volume MICROMEGAS TPC, operated using a gas mixture of high-purity argon and 5\% isobutane at 550 torr, with actinide targets mounted onto the central cathode~\cite{Heffner2014}. The fissionTPC has several advantages over standard ionization chambers, especially when it comes to providing insight into sources of systematic uncertainties. In general, charged particle tracks are reconstructed by combining the 2-dimensional $x-y$ information from the highly-segmented readout plane with the relative drift time of the charge deposited by the track~\cite{Nygren}. Further details on the design and operation of the \ftpc{} can be found in Ref.~\cite{Heffner2014}. 

\begin{figure}[ht]
 \includegraphics[width=1.\linewidth]{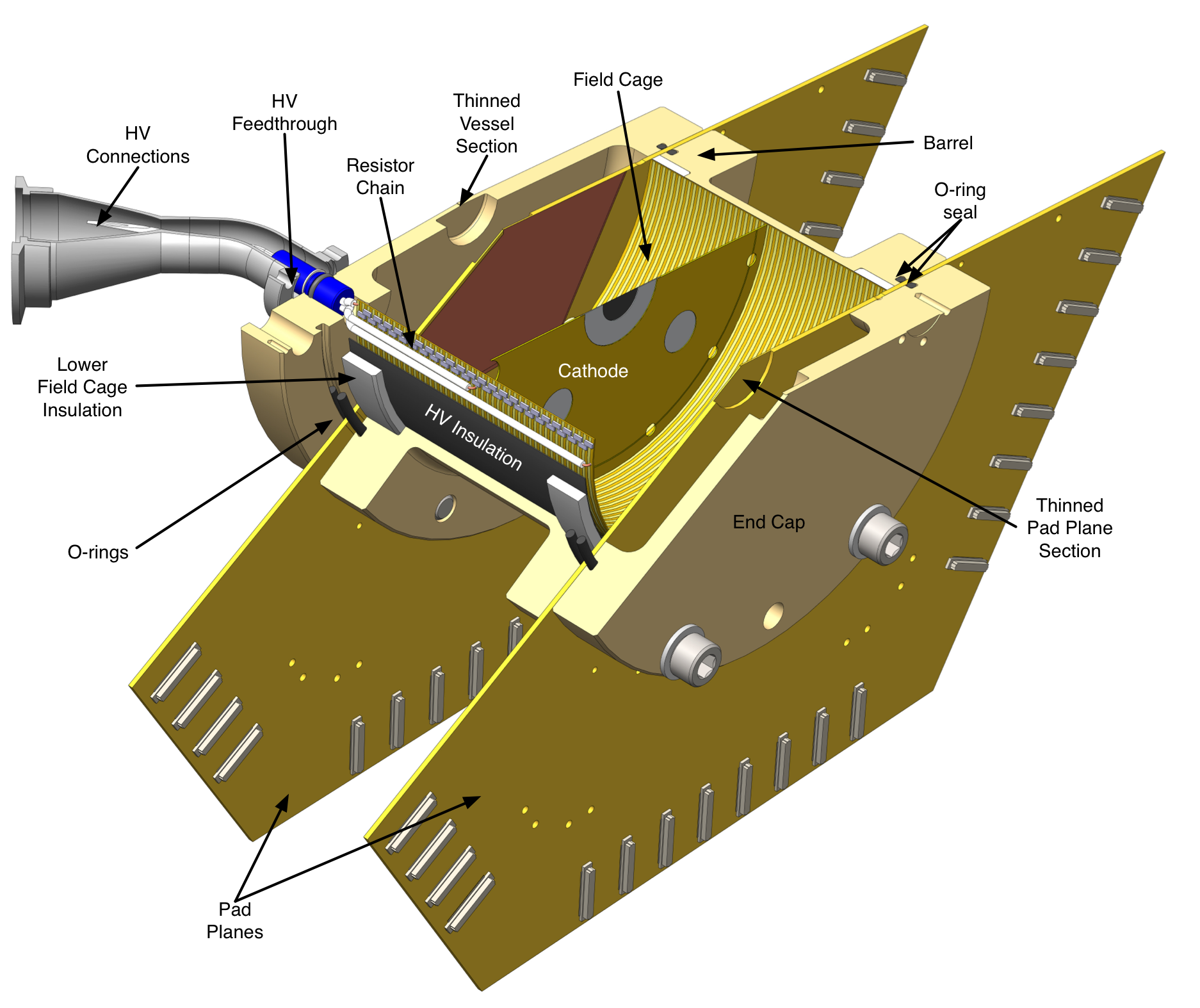}
 \includegraphics[width=1.\linewidth]{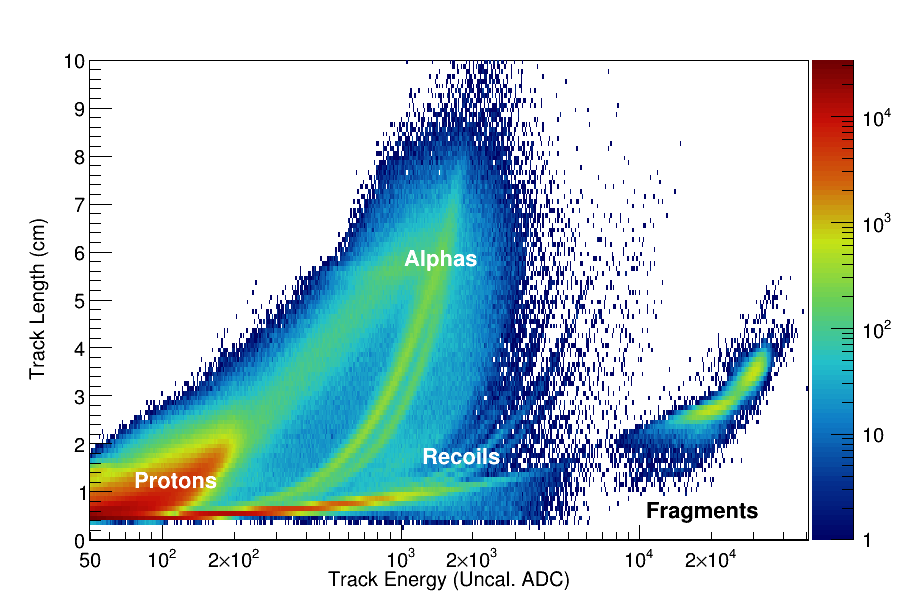}
 \caption{\label{fig:cutaway} (Top) A cutaway image of the \ftpc{}. (Bottom) Reconstructed data of charged-particle track length and energy provide sufficient information to separate the various particle interactions.}
\end{figure}

This high resolution particle tracking can help determine sample uniformity, beam profile, and other parameters which were often estimated rather than measured in earlier experiments. Additionally, the particle identification (PID) performance of the \ftpc{} enhances discrimination of fission fragments, alphas and other charged particles, shown in Fig.~\ref{fig:cutaway} (bottom). A complete description of the electronics can be found in Ref.~\cite{Heffner2013}, and further details on tracking data reconstruction for a fission cross-section ratio analysis are described in Refs.~\cite{Casperson2018,Snyder2021jor}.

In the cross-section ratio measurement published in 2021 the plutonium target was electroplated while the uranium target was vapor-deposited. Electroplating has the advantage of being very efficient in terms of the quantity of actinide lost, however making uniform targets can be challenging. On the other hand, vapor-deposition is inherently uniform but not as efficient. Vapor-deposition of plutonium is particularly challenging for two reasons; first, the chemistry required to get plutonium in a form suitable for vacuum volatilization, typically a fluoride, is difficult to accomplish. Second, the relatively high specific-activity of plutonium combined with the high dispersion of the vapor-deposition process present a substantial radiological hazard. For these reasons the vapor-deposition of a $^{239}$Pu target had not been accomplished, to our knowledge, for decades until recently.
The Radiochemistry Department at Oregon State University led by W. Loveland accomplished this by constructing a so-called dual-glovebox system. The dual-glovebox was essentially two gloveboxes attached with a pass-through chamber, better enabling radiological control by isolating high and low-contamination chambers~\cite{Loveland2016, Chemey:2022}.
Uniformity of the previous actinide targets was described in Sec.~\ref{sec:introduction} and is shown in Fig.~\ref{fig:vertDist}. The uniformity of the  plutonium target used for the results presented here is at the 10\% level, where uniformity is defined as the standard deviation of the distribution of normalized bin counts of radiograph data, as it will be shown in Sec.~\ref{sec:overlap}. 

\section{Data Analysis}
\label{sec:results}
The measured cross-section ratio is described by Eq.~\eqref{eqn:xsCalc}.  In this formulation $x$ denotes the unknown actinide, in this case \pu{}, whereas $s$ represents the reference standard actinide, in this case \u{},

\begin{align}
\label{eqn:xsCalc}
\frac{\sigma_x}{\sigma_s} =& \frac{N_s}{N_x}\frac{\omega_s}{\omega_x}\frac{\kappa_s}{\kappa_x}\frac{\epsilon_{ff}^s}{\epsilon_{ff}^x} \frac{\Phi_s}{\Phi_x} \frac{\sum_{XY}(\phi_{s,XY}\cdot n_{s,XY})}{\sum_{XY}(\phi_{x,XY}\cdot n_{x,XY})} \nonumber \\ 
~& \times \left(\frac{\left[(C_{ff}^x - C_r^x - C_\alpha^x) - C_{w}^x\right] \cdot C_b^x}{\left[(C_{ff}^s - C_r^s - C_\alpha^s) - C_{w}^s\right] \cdot C_b^s} \right).
\end{align}

Here $N$ denotes the number of target atoms, $\omega$ is the detector livetime, $\kappa$ accounts for the down-scatter and attenuation of neutrons from transport through the detector material and target backing, $\epsilon_{ff}$ denotes the fission fragment detection efficiency, $\Phi$ represents the neutron flux and $\sum_{XY}(\phi_{XY}\cdot n_{XY})$ is the spatial beam and target overlap term, which accounts for nonuniformity of the beam and target shape. The $C$ terms are detector counts, with $C_{ff}$ being the number of fission fragment candidate events which are then corrected for various background classes. $C_\alpha$ is the beam-uncorrelated, pile-up $\alpha$-decays misidentified as fission. $C_r$ is the beam-correlated background, or neutron-induced events misidentified as fission.  $C_w$ is the wraparound contribution, coming from fission events assigned an incorrect neutron energy.  Finally, $C_b$ is the correction for contamination of the targets with other fissile isotopes.  

The analysis procedure for extracting the $^{239}$P$(n,f)/^{235}$U$(n,f)$ cross-section ratio is intricate and involves many steps that have been outlined in great detail in Refs.~\cite{Snyder2021jor,Monterial2021}. Quantifying all of the experimentally accessible uncertainty contributions to the cross-section ratio as a function of neutron energy, including the full energy correlation matrix, rather than simply the uncertainties for each energy bin, is a primary aim of the NIFFTE Collaboration.  To accomplish this, multiple factors that contribute to the uncertainties have been addressed.


\subsection{Wraparound}
The LANSCE proton accelerator produces ``micro-pulses" or beam bunches spaced $\sim1.8\mu$s apart. The slower low-energy neutrons from one bunch may overlap with faster neutrons from later bunches. This results in contributions of the low-energy neutrons to the high-energy region of the time-of-flight distribution, which need to be corrected for.

\begin{figure}[ht]
 \includegraphics[width=1.\linewidth]{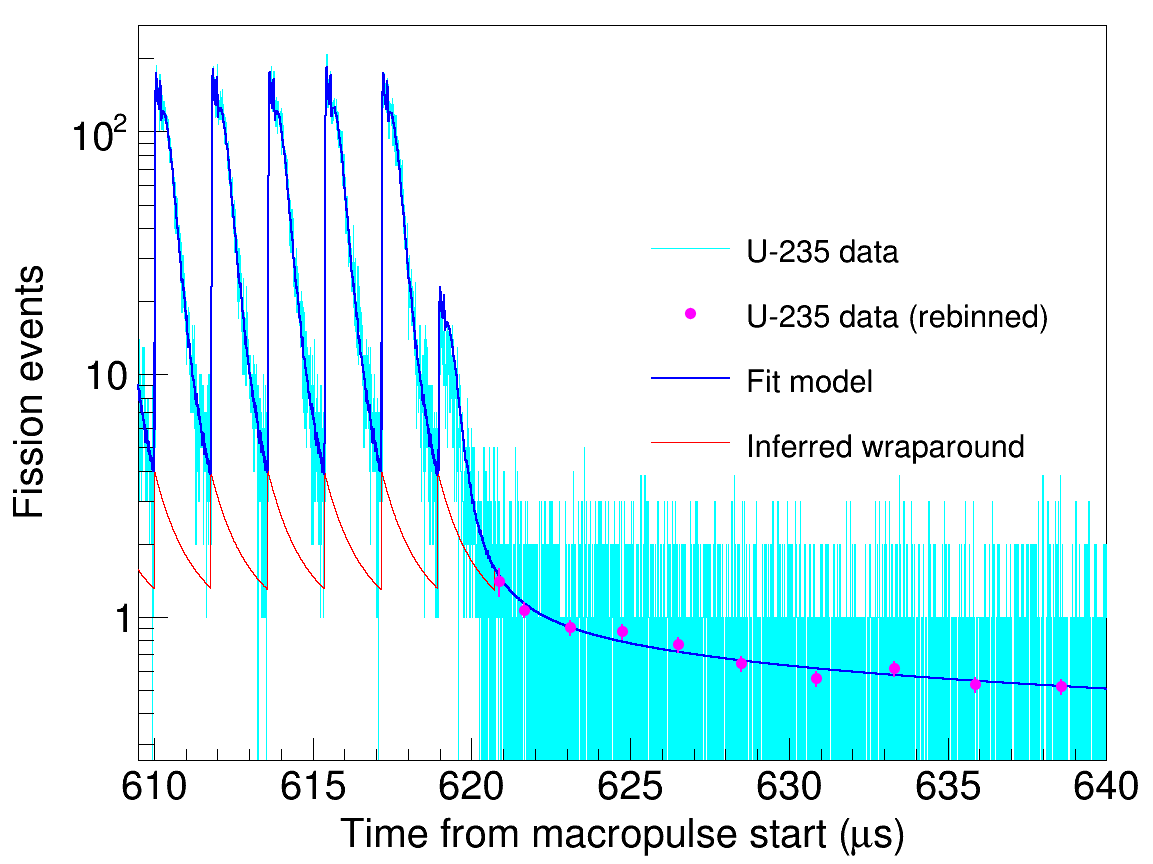}
 \includegraphics[width=1.\linewidth]{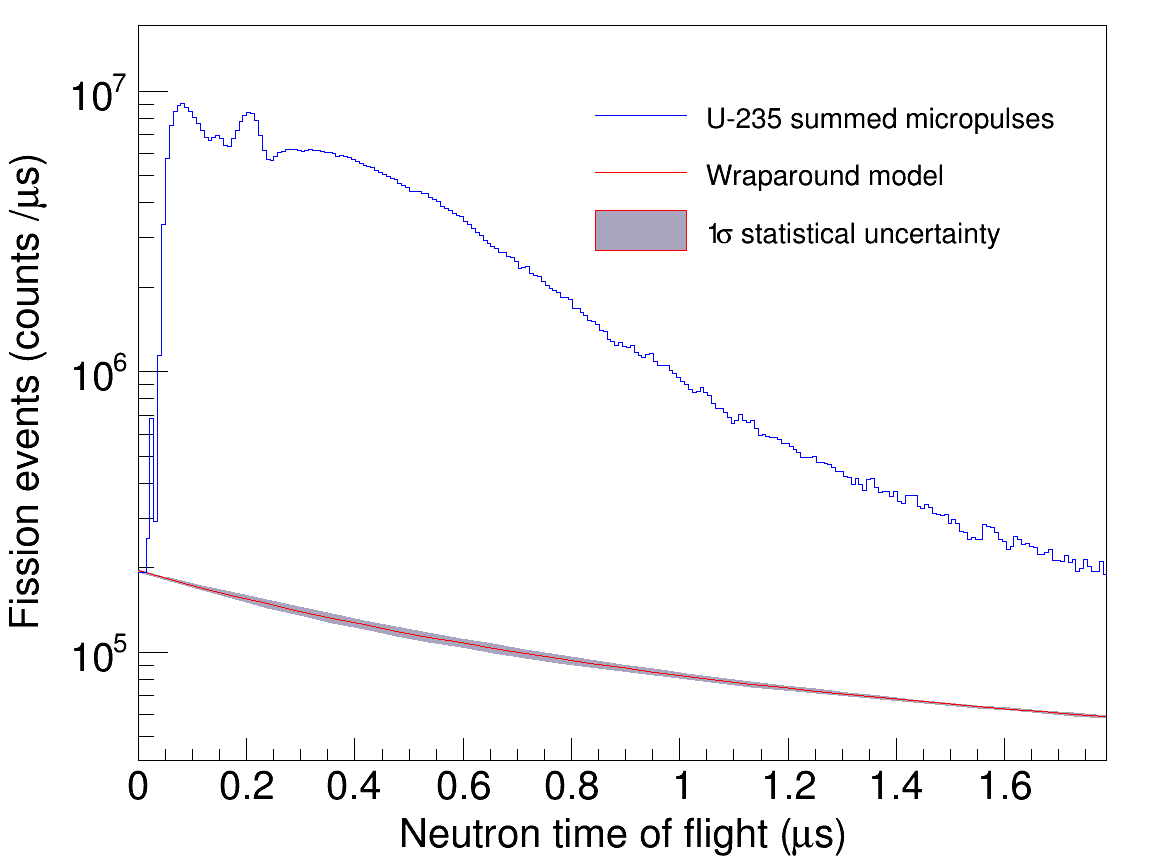}
 \caption{\label{fig:macpulse} Determination of the wraparound correction in the $^{235}$U-target side of the current dataset. Top: The nToF data (cyan), averaged in the low-energy tail region (magenta), are fit to determine the wraparound contribution (red line) to the nToF model (blue line). The wraparound model consists of a logarithmic spline following the distribution of prompt neutron data. Bottom: The nToF data and wraparound correction after all micropulses are combined. The band around the red line represents the uncertainty of the wraparound fit.}
\end{figure}

The recorded data continues $\sim70\mu$s beyond the last micro-pulse, and “unwrapping” the micro-pulse train structure reveals a long tail beyond the last micro-pulse, which can be fit in order to determine the wraparound contribution. Fig.~\ref{fig:macpulse} (top) depicts the micro-pulse structure within a macro-pulse, accompanied by the fit results that are used to correct for wraparound; the procedure is fully described in~\cite{Snyder2021jor} .

A logarithmic spline was used to characterize the wraparound contribution, as shown for the current $^{235}$U data in the top panel of Fig.~\ref{fig:macpulse}, since this was found to describe the low-energy tail of the nToF distribution well over large time scales. A total nToF distribution is constructed by combining the micro-pulses (see bottom panel of Fig.~\ref{fig:macpulse}). 
The uncertainty band of the wraparound correction is calculated with Monte Carlo variations based on the covariance matrix provided by the fitting algorithm.

\subsection{Efficiency}
\label{sec:Efficiency}

A phenomenological efficiency model was developed in order to determine the fission fragment detection efficiency $\epsilon_{ff}$ of the fissionTPC~\cite{Casperson2018}. The model used as input the detailed event-by-event information captured by the fissionTPC as a function of incident neutron energy. The efficiency model includes parameters to account for the fission fragment energy and angular distributions, including energy loss and transport in the target material, kinematic boost, and analysis selection cuts.  Full details of the efficiency model can be found in ~\cite{Casperson2018,Snyder2021jor}.
A brief summary is provided here, as well as a discussion of a minor difference in the handling of the efficiency model between the current and previous measurement.

The fissionTPC employs a ``level-2'' trigger masking scheme to suppress the otherwise unsustainable data rate that would be induced by the $^{239}$Pu $\alpha$-decay. The trigger mask and its performance validation are described in \cite{Snyder2021jor}. 
The fast cathode amplifier gain for this new measurement was slightly lower than in the previous measurement, which resulted in the level-2 threshold cutting the low-energy tail of the fission fragment distribution.
This portion of the distribution is usually removed by the selection cut and corrected for with the efficiency model, however the tail is needed to fine-tune the model. The level-2 system was designed in anticipation of this eventuality and is disabled automatically, at regular intervals (every 8$^{th}$ macro-pulse in this case) throughout a data collection run.  This provides a subset of data, evenly distributed throughout the full data set, that can be used to tune the efficiency model. In our previous measurement the cathode threshold was below the level-2 mask threshold and the model could be tuned directly on the full dataset.

\begin{figure}[ht]
 \includegraphics[width=1.\linewidth]{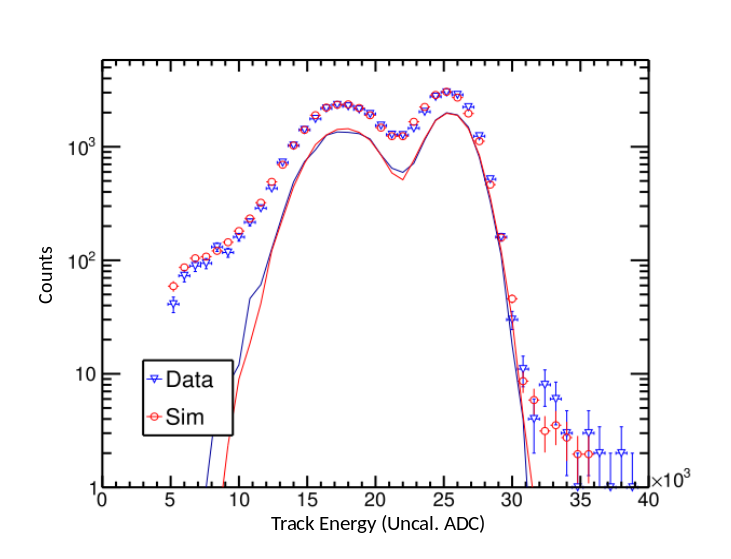}
 \includegraphics[width=1.\linewidth]{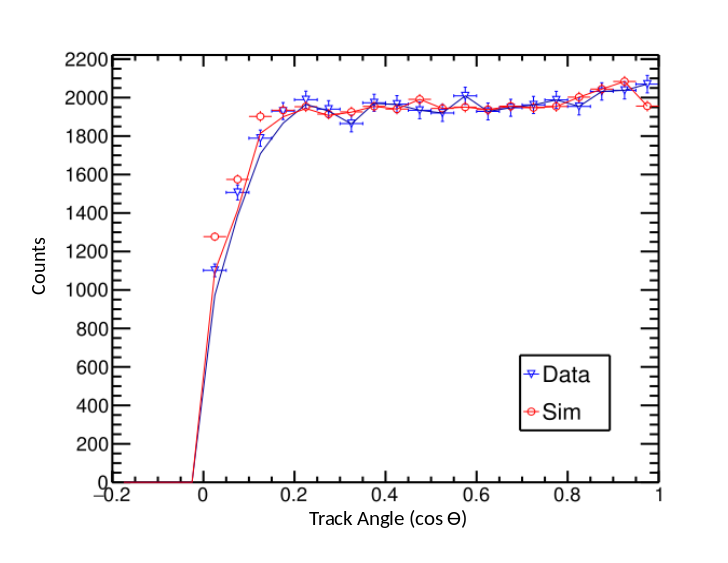}
 \caption{\label{fig:m0projections} 
 Projections of fission fragment energy (top) and angle distributions (bottom) comparing data and simulation for the low threshold level-2 data-subset that was used to optimize fit parameters for the entire data suite. The solid lines represent the counts after the selections cuts on $\cos(\theta)$ (top) and track energy (bottom) detailed in \ref{app:frag_selection}.
 }
\end{figure}

The result of the efficiency model fit to the level-2 data-subset are shown in Fig.~\ref{fig:m0projections}.  
Best-fit parameters that result from the $\chi^2$ minimization, taking into account the fragment transport (energy loss) and angular distributions, were extracted.
For the full dataset,
evaluation of the fission fragment detection efficiency was carried out by histogramming each isotope for every neutron energy bin, and applying the best fit parameters found by tuning to the level-2 data-subset. The expected fission fragment distributions as a function of incident-neutron energy were then calculated. Fission fragment transport and anisotropy models provide the fraction of particles entering the fissionTPC that would pass analysis selection cuts. 

\begin{figure}[ht]
 \includegraphics[width=1.\linewidth]{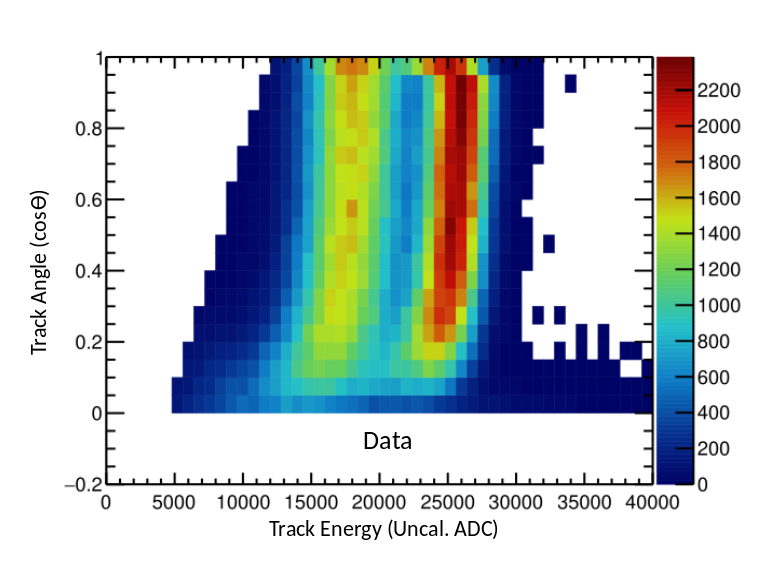}
 \includegraphics[width=1.\linewidth]{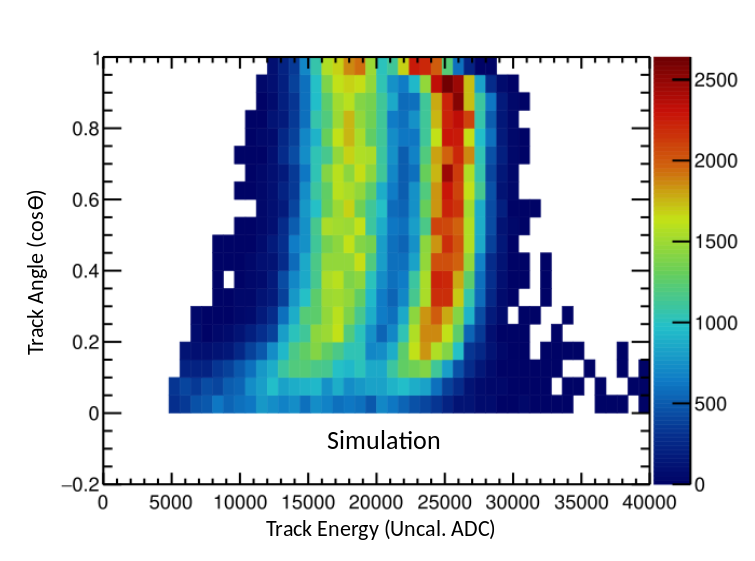}
 \caption{\label{fig:simData} Fission fragment angle of emission vs. energy for data (top) and simulation (bottom) from the $^{235}$U side of the target induced by neutrons with energies between 10.0 and 10.6 MeV. The two bands represent the heavy and light fission fragment distributions.}
\end{figure}

\begin{figure}[ht]
 \includegraphics[width=1.\linewidth]{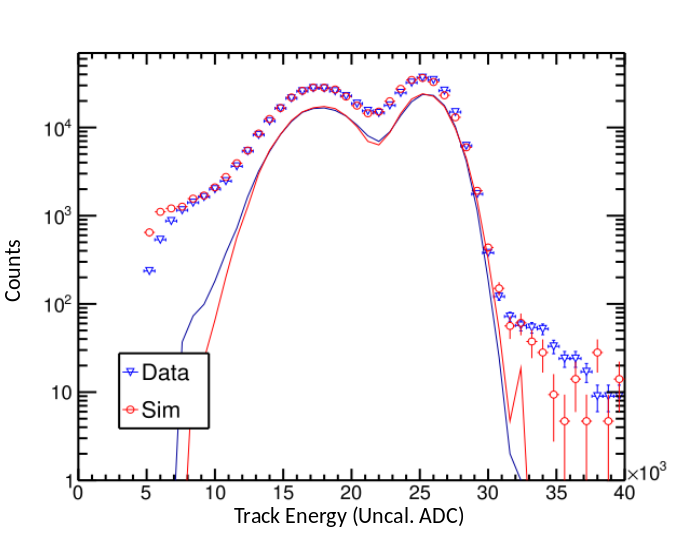}
 \includegraphics[width=1.\linewidth]{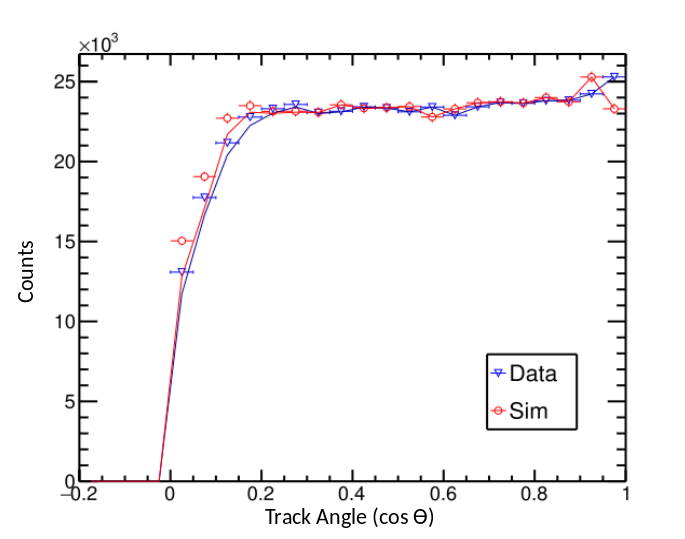}
 \caption{\label{fig:projections} 
 Projections of fission fragment energy (top) and angle distributions (bottom) comparing data and simulation for the full dataset. The small discrepancy at low energy is a result of the level-2 trigger threshold discussed in the text. The solid lines represent the counts after the selections cuts on $\cos(\theta)$ (top) and track energy (bottom) detailed in App.~\ref{app:frag_selection}.
 }
\end{figure}

\begin{figure}[ht]
\includegraphics[width=1.\linewidth]{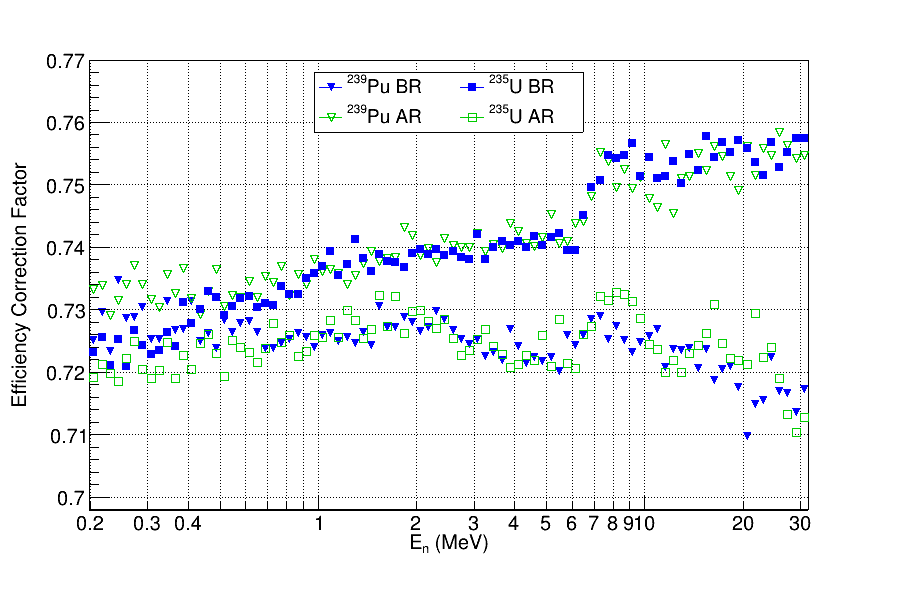} 
 \caption{\label{fig:effic} Calculated fission fragment detection efficiency for the $^{239}$Pu and $^{235}$U targets before (BR) and after (AR) 180 degree rotation of the fissionTPC. The difference in the efficiency curves for the two orientations primarily reflects kinematic and anisotropy effects.}
\end{figure}

Results for the full dataset on a 2-dimensional parameter space of track energy and detection angle for both data and simulation are shown in Fig.~\ref{fig:simData}. The two bands represent the heavy and light fission fragment distributions. At track angles approaching $\cos(\theta)=1$ the effects of electronics saturation are observed, particularly for the higher energy fragment band. The bending of the two bands towards lower energy, observed for track angles approaching $\cos(\theta)=0$, results from energy loss in the target material. 
Fig.~\ref{fig:projections} shows projections of the full 2-D distributions of Fig.~\ref{fig:simData} along each axis, demonstrating the level of agreement between simulation and data. The tracks originate from fission events induced in the $^{235}$U side of the target by neutrons with energies between 10.0 and 10.6 MeV. 
The efficiency results as a function of neutron energy, before and after the fissionTPC was rotated $180^\circ$, are shown in Fig.~\ref{fig:effic}.

The orientation of the fissionTPC relative to the beam direction, which has significant impact on the angular distribution of the fission fragments, and therefore on the efficiency, has been extensively discussed in Ref.~\cite{Snyder2021jor}. 
The efficiency values are lower for the current data compared to the previous $^{239}$Pu/$^{235}$U target reported in~\cite{Snyder2021jor}, since a stricter energy cut has been applied here that reduces the overall acceptance.
This higher cut was levied in order to match the higher relative cathode threshold (lower cathode gain) as previously discussed. Details of the selection cuts are provided in \ref{app:frag_selection}.

The general trend in the efficiency as a function of energy is a result of the kinematic boost from neutron momentum transfer. Fission fragments detected in the downstream volume relative to the neutron beam are boosted in the forward direction by momentum transfer, away from the target plane, thereby increasing the number of fragments entering that volume, and vice versa for the upstream volume. The characteristic energy-dependent structure of the efficiencies is associated with fission anisotropy, which changes rapidly with the onset of second-chance fission around 6 MeV.


\subsection{Beam and Target Overlap}
\label{sec:overlap}
Experiments to measure fission cross-section ratios are typically arranged so that the two targets share the same beam, to ensure that the neutron flux term in ratio cancels out.  If, however, \emph{both} the beam and the targets have spatial nonuniformities, a correction is needed.  The correction term is denoted as $\sum_{XY}(\phi_{XY}\cdot n_{XY})$ in Eq.~\ref{eqn:xsCalc}. The implementation of this overlap correction with fissionTPC data is detailed in Ref.~\cite{Snyder2021jor}. 

For this analysis, several confounding factors were discovered during analysis that necessitated careful treatment and application of various unforeseen corrections.  These include a difference in the sizes of the uranium and plutonium targets, 
a shift in mask placement relative to target backing
between layers of the vapor-deposited plutonium target, and differences in gain settings applied during data collection with beam in this experiment compared to the 2021 published fissionTPC result.  The impacts of these factors and the corrections developed to address them are described in this and subsequent sections.  
The fact that these issues could be observed in detail and correction factors could be quantified reinforces the robustness of the fissionTPC for identifying sources of uncertainty in cross-section ratio measurements that might otherwise go unnoticed.

For consistency with the previous measurement~\cite{Snyder2021jor}, the beam and collimation remained the same, resulting in the same level of beam non-uniformity.  While the targets for this measurement were both uniform at the 10\% level, due to a miscommunication with the manufacturer, the diameters of the $^{235}$U and $^{239}$Pu targets were different, with the plutonium target having a radius of 1 cm, while the uranium target radius was 0.75 cm.   Although the goal with this re-measurement was to eliminate the need for a beam-target overlap correction, the differently sized targets, when convolved with the nonuniform beam, required one.  As will be shown, the consistency of the corrected results from both the previous and current analyses validates the overlap correction procedure, particularly given that the size and shape of the needed corrections are substantially different than in the previous analysis.

\begin{figure}[ht]
 \includegraphics[width=1.\linewidth]{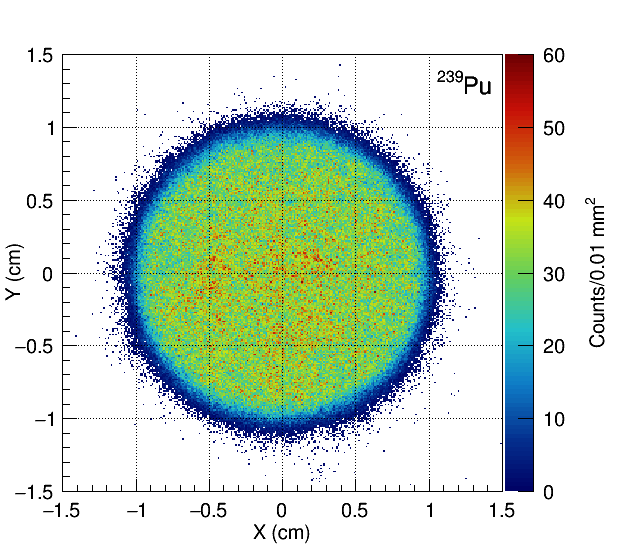}
 \includegraphics[width=1.\linewidth]{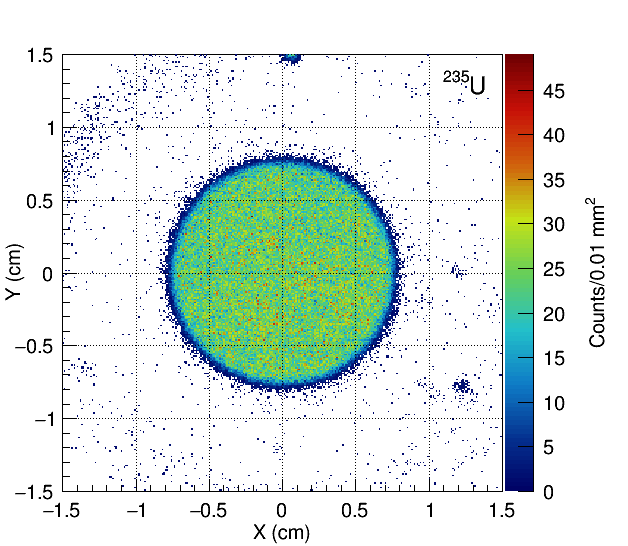}
 \caption{\label{fig:radtgts} $\alpha$-track vertices of the vapor-deposited $^{239}$Pu (top) and $^{235}$U (bottom) targets.}
\end{figure}

When the detector occupancy is large, as is the case for the high $\alpha$-decay activity $^{239}$Pu target side, distortions in the electric field cause a so-called ``space-charge effect’’, as described in the previous work~\cite{Snyder2021jor}. The small, but measurable, distortion in the electric field caused by positive ion back-flow from the MICROMEGAS is radially symmetric and pulls the drifting electron clouds toward the center of the chamber.  This results in the reconstructed track vertices for both $\alpha$-tracks and fission fragments also being pulled toward the center, and the resulting radiograph image of the target appears to have a smaller diameter than in reality. 

A procedure for correcting this effect was developed that relies on comparison of the reconstructed track vertices with a photograph of the target.  The position of the reconstructed track vertices are scaled to match the radius of the target from the photo. Since no space-charge effect is expected on the uranium side, due to much lower $\alpha$-decay activity, the reconstructed track vertices are expected to match the radius in the photo of the uranium target.  This was confirmed in both this and the previously published analysis, validating the approach for correcting for space-charge on the plutonium side. Unfortunately, the misaligned vapor-deposited layers of the plutonium target used for this analysis and the difference in gain compared to the previous work, required a more extensive treatment that is described in detail below.

\begin{figure}[ht]
 \includegraphics[width=0.95\linewidth]{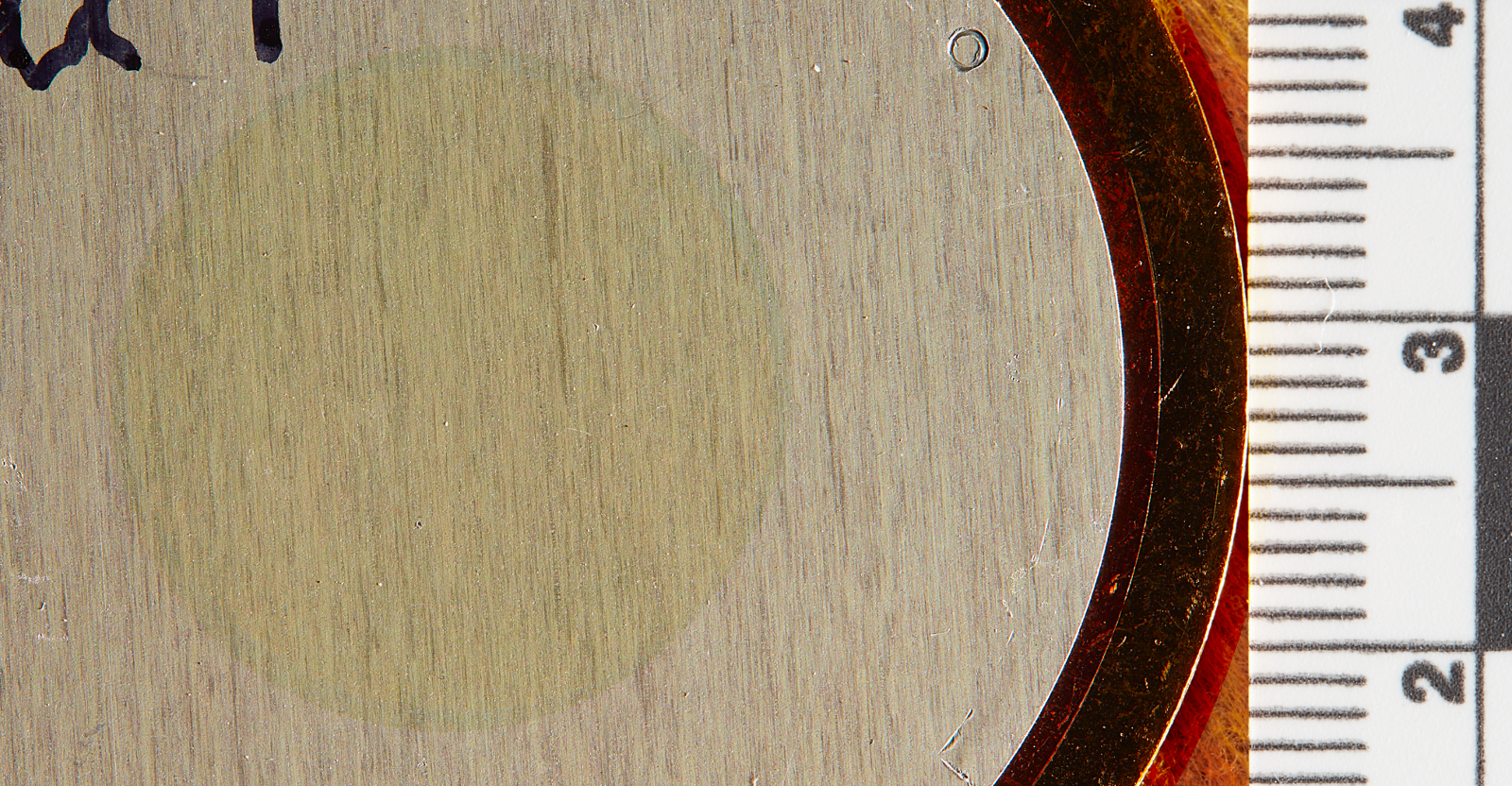}
 \includegraphics[width=0.95\linewidth]{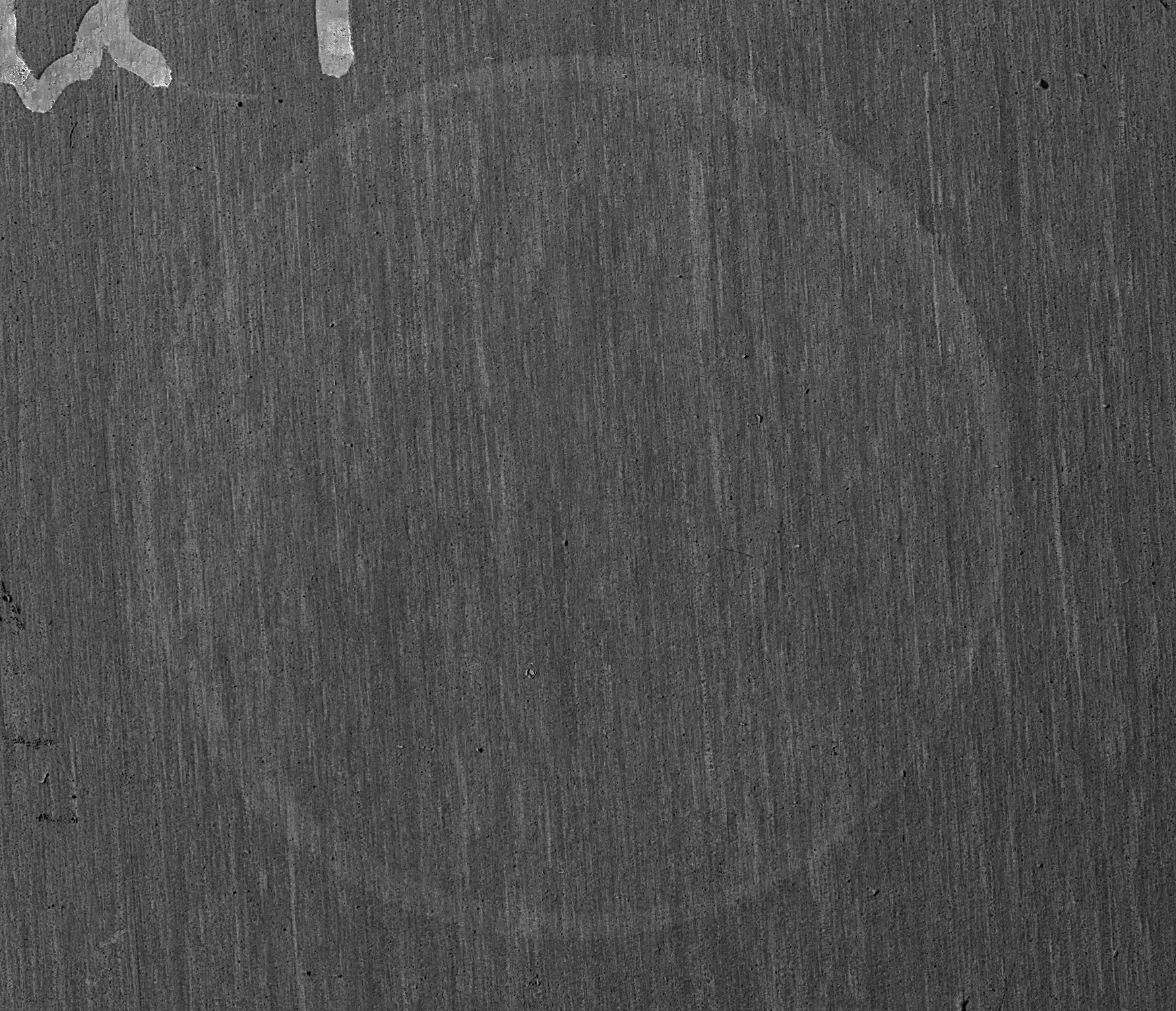}
 \caption{\label{fig:photo} The top panel shows a photograph of the vapor deposited plutonium on an aluminum backing along with a reference ruler.  The evidence of the misaligned deposits appears as a slight change in color along the edge of the upper right and lower left quadrants. The lower panel shows a zoomed in image of the deposit with color inversion and desaturation in an attempt to better highlight the color changes.}
\end{figure}

The vapor deposition of the plutonium target was applied in two separate depositions. In between depositions the mask was inadvertently moved by a small amount, which resulted in two misaligned circular deposits of plutonium.  This misalignment in the deposition is apparent in a photograph of the target and was measured to be 0.726 mm (see Fig.~\ref{fig:photo}).  The offset in the deposits was also evident in the radiograph data of the target. Since the offset is small compared to the pointing resolution of the fissionTPC tracking, it appears as a change in pointing resolution around the circumference of the target rather than as two distinct target edges (see Fig.~\ref{fig:radPhiFit}).  The consequence of this offset is that a single value for the target radius cannot be extracted from the photo and the radiograph data. Rather, a more extensive analysis of the target radiograph data was required, which is described in the following Sec.~\ref{sec:offset}.

The second challenge stems from the relatively lower gain of the beam data compared to that of the previous measurement.  While this lower gain reduced the amount of saturation that was observed in the fission fragments and thereby improved their tracking, it also degraded the quality of the pointing resolution of radiograph data collected at the same gain, to the extent that it was not useful for the overlap analysis.  Additional radiograph data, taken at higher gain, was also collected but this change in gain also changed the number of positive ions generated in the MICROMEGAS gain stage and therefore the amount of positive ion back-flow.  In turn, the space charge observed in the higher gain radiograph data differs from that of the beam data with fission fragments present.  Further radiograph data was therefore collected and analyzed, with varying gain settings, in order to confirm the expected trend of the space charge as a function of gain.  This analysis is described in Sec.~\ref{sec:gain}.

\subsubsection{Offset Target Deposit Analysis}
\label{sec:offset}  

\begin{figure}[ht]
    \centering
    \includegraphics[width=1.\linewidth]{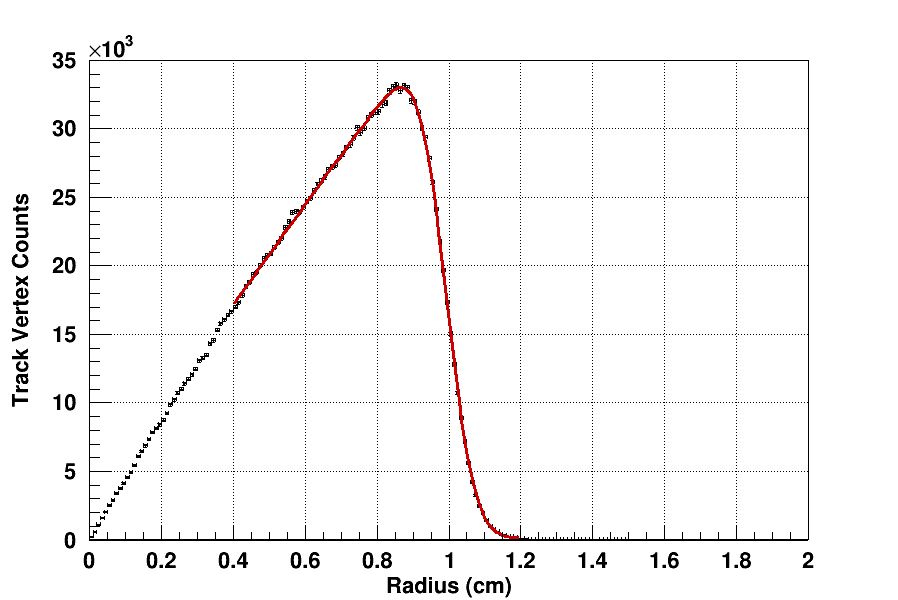}
    \caption{The radius of the plutonium $\alpha$-track vertices from the radiograph data. The edge is fit with the error function of Eq.~\ref{eqn:erf} to determine radius and pointing resolution.}
    \label{fig:radFit}
\end{figure}

\begin{figure}[ht]
    \centering
    \includegraphics[width=1.\linewidth]{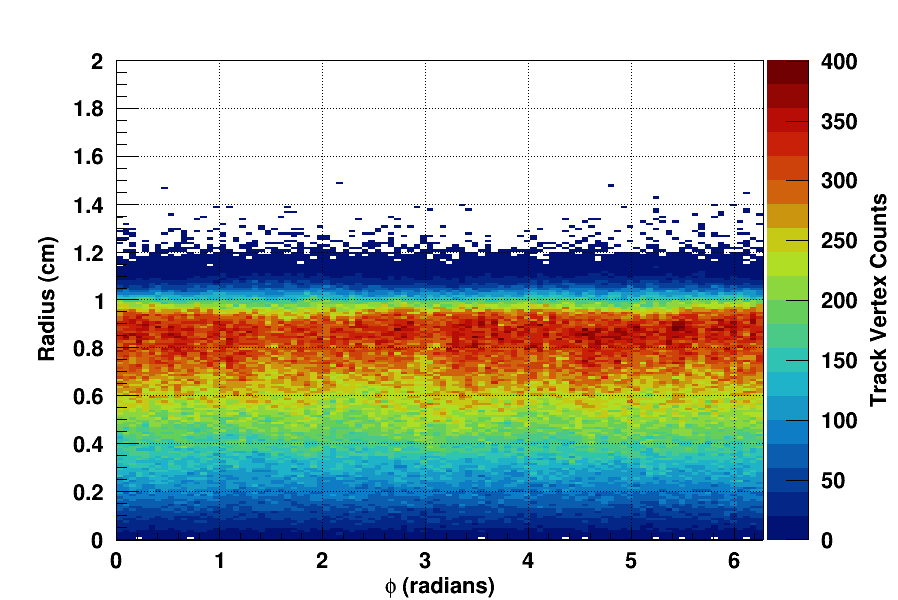}
    \caption{Radiograph data $\alpha$-track radius vs.~the angle about the target center, $\phi$.}
    \label{fig:radPhi}
\end{figure}

\begin{figure}[ht]
    \centering
    \includegraphics[width=1.\linewidth]{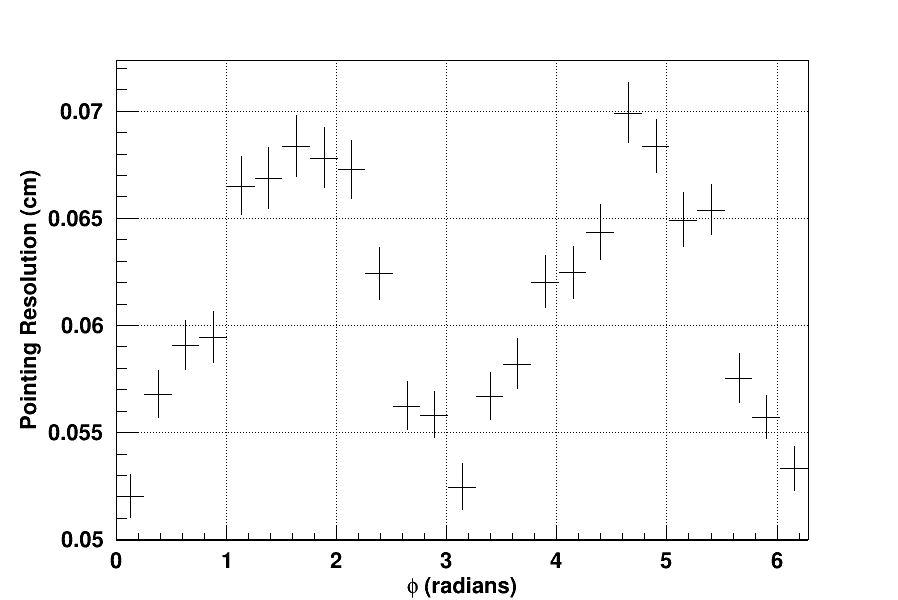}
    \caption{The pointing resolution from the fit of each $\phi$ bin of Fig.~\ref{fig:radPhi} with Eq.~\ref{eqn:erf}. This shows the apparent change in pointing resolution along the circumference of the target, which is a result of the misaligned deposits.}
    \label{fig:radPhiFit}
\end{figure}

The radius and pointing resolution of the radiograph data were obtained by plotting the radii reconstructed from the $\alpha$-track vertex and fitting the data with an error function of the form given by Eq.~\ref{eqn:erf}, where $p_1$ is the target radius and $p_2$ is the pointing resolution. A fit result is shown in Fig.~\ref{fig:radFit}.  
\begin{equation}
    p_0\cdot x^{p_3}\left(1 - \text{erf}\frac{x-p_1}{\sqrt{2}p_2}\right)+p_4
    \label{eqn:erf}
\end{equation}
By fitting the histogram of track radius vs.~the angle about the target center, $\phi$, also with Eq.~\ref{eqn:erf} and shown in Fig.~\ref{fig:radPhi}, any change in target radius and track pointing resolution along the circumference of the target deposit can be observed in Fig.~\ref{fig:radPhiFit}.  
Since the misalignment was 0.726 mm compared to the pointing resolution of $\sim$ 0.5 mm, rather than observing a shelf-like feature in the plot of target radius, the reconstructed average radius shows what appears to be a change in pointing resolution. 
The pointing resolution is taken to be, at most, the minimum shown in Fig.~\ref{fig:radPhiFit}, which would be the crossing point of the circumferences of the two misaligned deposits.

In order to determine the target radius space charge correction, a simulation of track vertices from the two misaligned, uniform deposits was performed.  The size and offset of the deposits in the simulation were informed by the analysis of the photograph and the pointing resolution was set to be the minimum from Fig.~\ref{fig:radPhiFit}.  The simulation then varied the average center of the deposits, relative to the pad plane center, the rotational orientation of the major axis of the deposit misalignment, and the radial space charge effect.  A $\chi^2$ comparison was performed between the data and simulation histograms of the track vertex radius in \emph{each} $\phi$ bin of Fig.~\ref{fig:radPhi}. The $\chi^2$ for each $\phi$ bin was then added to provide a single $\chi^2$ value for each set of simulation conditions.  The features of the fitted target radius vs.~$\phi$ histogram were dominated by the simulated $x,y$ center, while the pointing resolution vs.~$\phi$ histogram features were dominated by the misalignment's rotational orientation, as expected. 

\begin{figure}[ht]
    \centering
    \includegraphics[width=0.95\linewidth]{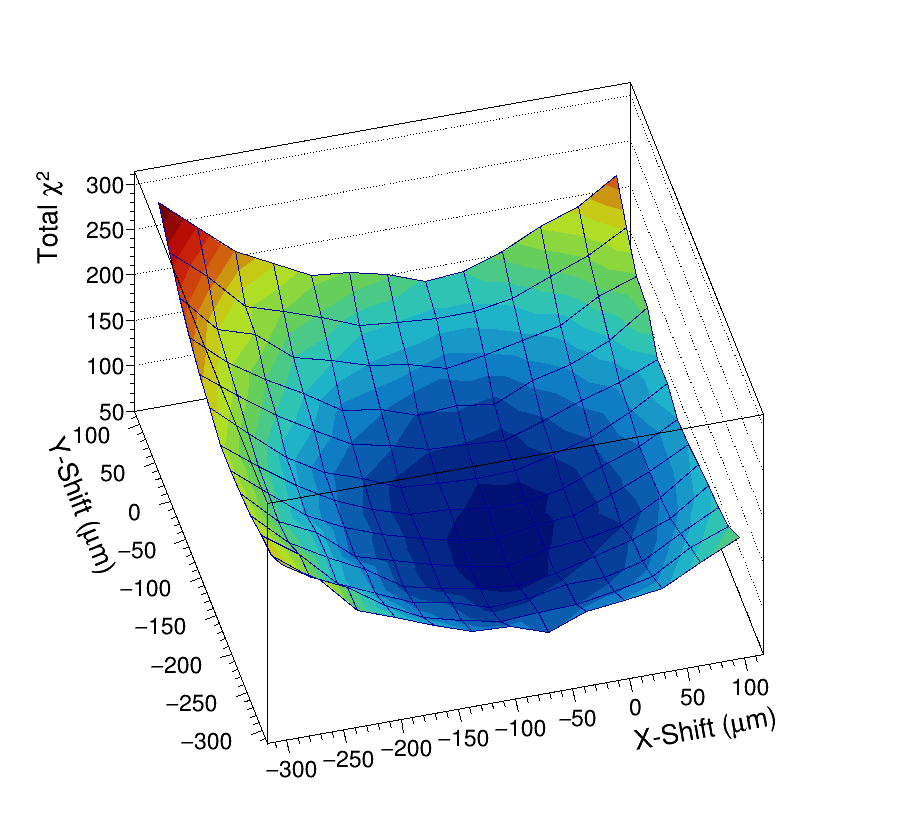}
    \includegraphics[width=0.95\linewidth]{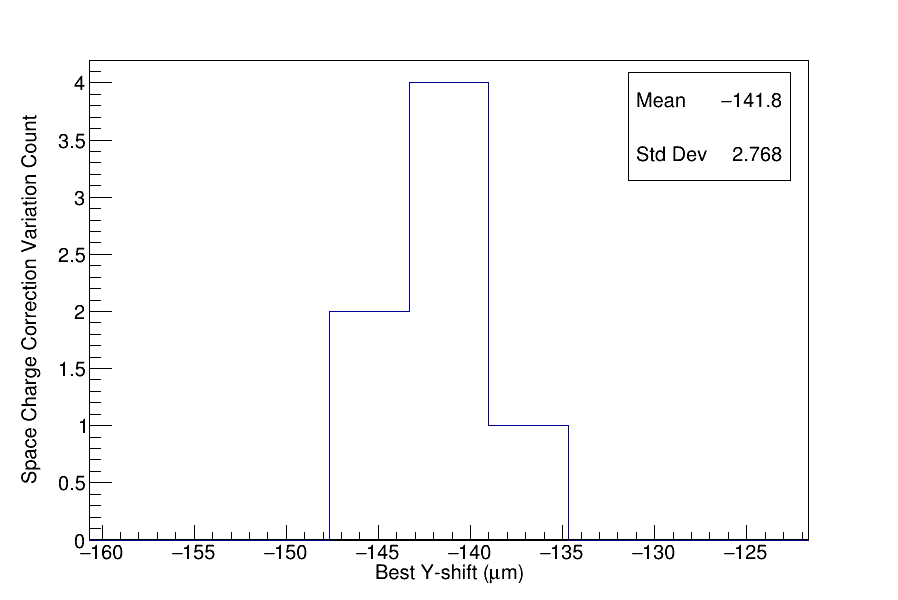}
    \caption{The top panel shows the distribution of total $\chi^2$ comparison between data and simulation for various shifts in the relative (x,y) center of the simulated target while the other simulation parameters are held fixed.  Note that the $\chi^2$ shown here is the sum of 50 comparisons from each $\phi$ bin of Fig~\ref{fig:radPhi}, the $\chi^2$ in each bin is of order 1. The lower panel shows the result of a parabolic fit along the y-axis of the top panel to find the minimum $\chi^2$, as the space charge correction parameter is varied.  The best (x,y) shift as determined by the simulation is largely independent of the other parameters within $\sim$3 $\mu m$.}
    \label{fig:minChi}
\end{figure}

\begin{figure}[ht]
    \centering
    \includegraphics[width=1.\linewidth]{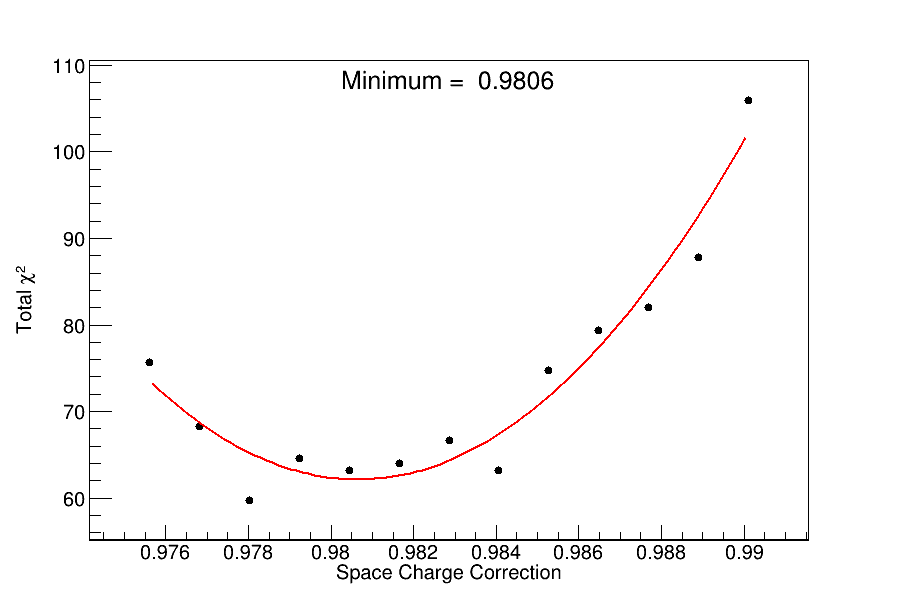}
    \caption{The $\chi^2$ distribution as the space charge parameter is varied while all other parameters are held fixed at their best values.}
    \label{fig:bestSpace}
\end{figure}

The simulation parameters are uncorrelated, such that the $\chi^2$ distributions can be fit with simple parabolas to determine the best $x,y$ center, misalignment orientation, and most importantly the space charge correction independently.  This assertion is supported by Fig.~\ref{fig:minChi}, the top panel of which shows the $\chi^2$ as the $x,y$ center of the target simulation is varied while space charge and misalignment orientation are held fixed. The lower panel plots the minimum of a parabolic fit along the y-axis of the top panel as the space charge parameter is varied.  In this case the best fit y-offset (or x-offset) is reproduced within a reasonable standard deviation of $\sim$3 $\mu m$ independent of the space charge parameter.  Similar behavior is observed as the misalignment orientation is varied. The $\chi^2$ distribution vs.~space charge parameter can be fit to a parabola while holding the center offset and orientation parameters fixed at their best values to determine the best space charge correction for the data, which is shown in Fig.~\ref{fig:bestSpace}. 

\subsubsection{Variable Gain Space Charge Analysis}
\label{sec:gain}
The beam data collected for the fission cross-section measurement was at a lower gain than in the previous measurement.  While the fission fragments, which have a much greater and denser specific energy loss compared to $\alpha$-particles, were tracked well in this configuration, the beginning of $\alpha$-tracks were near or below the fissionTPC pad thresholds.  The vertex pointing resolution of $\alpha$-tracks at this lower gain setting was not adequate enough for the overlap correction.  While increasing the gain would improve the radiograph data, it would also increase the relative size of the space charge correction, since both the fragment data and radiograph data are needed to extract the overlap correction.  The fission fragments alone cannot be used to extract the space charge, since the beam nonuniformity results in a fission fragment vertex distribution that is also nonuniform and not well represented by the fit function of Eq.~\ref{eqn:erf}.

\begin{figure}[ht]
    \centering
    \includegraphics[width=1.\linewidth]{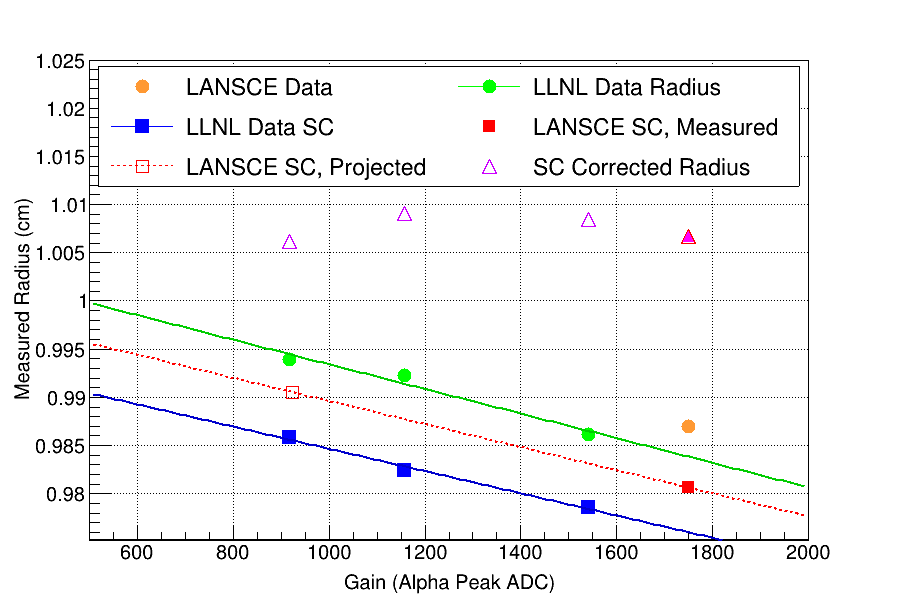}
    \caption{The measured target radius for data collected at LLNL as the gain is varied. Also shown is the space charge (SC) effect determined by a comparison between the radiograph data and the target photo as described in Sec.~\ref{sec:offset}. Note that the vertical axis is in $cm$ for the radius but the value of the space charge correction is fractional, i.e. the measured target radius is reduced by said fraction.}
    \label{fig:SpaceGain}
\end{figure}

The fissionTPC was therefore reassembled at LLNL after beam data collection at LANSCE was complete and radiograph data at various gains were collected to determine how the space charge correction depends on gain.  Fig.~\ref{fig:SpaceGain} shows a plot of the measured target radius, as determined by a fit of Eq.~\ref{eqn:erf}, vs.~gain for the radiograph data collected at LLNL and also the data point from LANSCE.  Also shown is the space charge correction (SC) as determined by the method described in Sec.~\ref{sec:offset}. The exact gain of the original radiograph data collected at LANSCE could not be replicated, as an instability developed in the MICROMEGAS that prevented higher gain settings.  This is not unusual for MICROMEGAS, as a discharge can result in permanent damage that limits the maximum achievable gain. For this analysis the tracking resolution at lower gains was improved by minimizing the thresholds.  A trend is clearly visible in the data.  A line projected through the LLNL data to the higher gain of the original radiograph data shows a $<$0.5\% difference.  This small difference is expected and can result from differences in experimental conditions between the lab at LLNL and at LANSCE.  

To determine the space charge correction for fission fragments, the trend in the space charge as a function of gain was assumed to be the same in both the LLNL and LANSCE data.  The trend line was linearly projected back along the same slope as the LLNL data with the intercept adjusted to match the LANSCE data.  Several different fissionTPC configurations were varied in an attempt to change the magnitude of the space charge so that it better matched the LANSCE data and to validate the stability of the trend as a function of gain.  The alterations included swapping the pad plane (and therefore the MICROMEGAS), doubling the isobutane content from 5\% to 10\%, dropping the gas pressure by 10\%, and adjusting field cage voltages. The general trend of the space charge remained the same for all of these configurations.

Two changes in particular had a measurable impact on the magnitude of the space charge. Decreasing the cathode voltage by $\sim$20\% increased the effect of the space charge by $\sim$ 1\%. This is likely due to a change in mesh transparency, which is the ability of the MICROMEGAS to capture the positive ions from the gain stage, given that mesh transparency is affected by the ratio of the field strengths in the drift region and the amplification region \cite{Nikolopoulos:2011}.  In the fissionTPC the bottom of the field cage (FCB), the anode side, is typically not set to ground but is adjusted to match the voltage on the MICROMEGAS to avoid field distortions~\cite{Heffner2014}. Adjusting the FCB voltage down decreased the space charge effect. This likely caused a minor compensating field distortion which opposed the space charge effect. This assertion is supported by the fact that adjusting the FCB voltage down had a minor effect on the reconstructed radius on the uranium side, where no significant space charge is present.  
Under this condition the LLNL and LANSCE data were in best agreement, indicating that equipment used to set the FCB voltage may have been set in the same condition during data taking at LANSCE.  As stated earlier, the slope of the trend in the space charge was similar for all operating conditions of the fissionTPC that were tested, and it is this trend that was used to calculate the space charge effect on the fission fragments. 

\begin{figure}[ht]
    \centering
    \includegraphics[width=1.\linewidth]{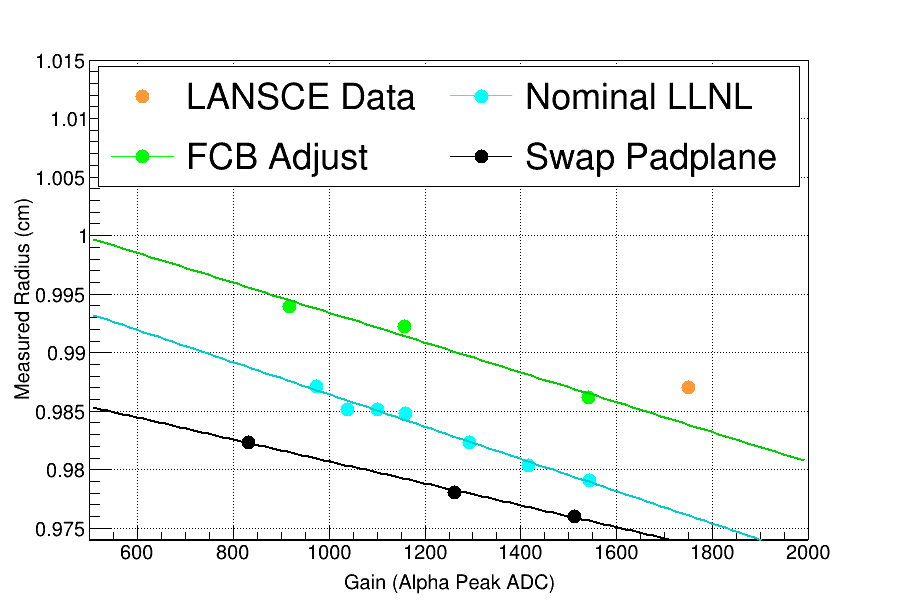}
    \caption{The measured target radius vs.~gain for data collected at LLNL as the fissionTPC operating configuration was varied. The ``FCB Adjust'' data is also the ``LLNL Radius Data'' shown in Fig.~\ref{fig:SpaceGain}}
    \label{fig:Changes}
\end{figure}

Careful consideration of the potential systematic uncertainty in the space charge correction is clearly needed.  One could consider taking the systematic difference between the LLNL and LANSCE data shown in Fig.~\ref{fig:SpaceGain} as an estimate of the potential uncertainty.  However, this would likely be an overestimate, because the MICROMEGAS operating conditions were different, as explained above.  As various operating conditions were altered, the shifts in the space charge were distinct and measurable. Fig.~\ref{fig:Changes} shows the measured track radius vs.~gain for three configurations of the fissionTPC which include changing the pad plane and the FCB drift field voltage.

There are distinct shifts in the measured radius for each particular data set, and while the slope of the trends with gain are similar, there are differences. It is the uncertainty in the slope that is of concern, since the calculation of the overlap correction relies on the slope of the radius vs.~gain to project back to the lower gain settings of the beam data. 
The mean of these measured slopes was therefore used to project linearly to lower gain, with the standard deviation as the uncertainty.  A Monte Carlo variational analysis was done to produce a systematic uncertainty for the overlap correction of 0.343\%. 
Changes to the gas pressure/mixture or cathode voltages were not consideded in this calculation, as any differences between LLNL and LANSCE operating conditions of those parameters, significant enough to impact the space charge, would have had other impacts on the data that would have been immediately clear, such as drift velocity and $\alpha$-track lengths.

\begin{figure}[ht]
    \centering
    \includegraphics[width=1.\linewidth]{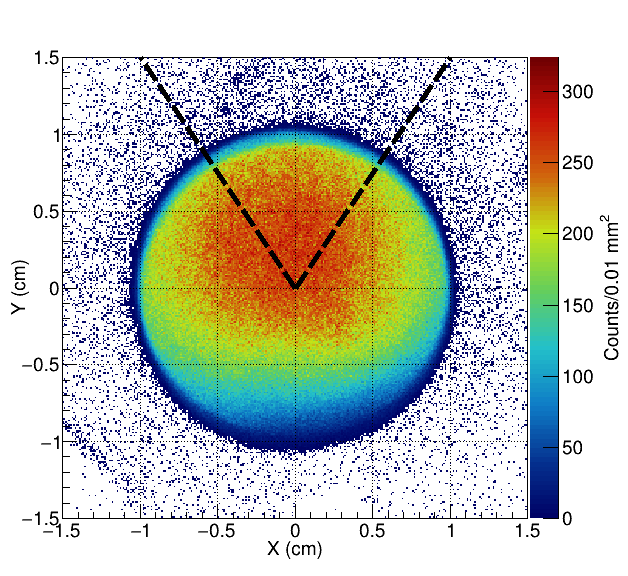}
    \caption{Fission fragment track vertices. The dashed line represents a region where the target edge is well defined and could be fit by Eq.~\ref{eqn:erf}.}
    \label{fig:FF_Tracks}
\end{figure}

The nonuniform neutron beam results in a nonuniform distribution of fission fragment track vertices as shown in Fig.~\ref{fig:FF_Tracks}.  As previously discussed, the target edge represented by fission fragment tracks is not well represented by the fit function of Eq.~\ref{eqn:erf}, and the data could not be used to directly determine the space charge.  There is a portion of the target edge that is well illuminated by a relatively uniform region of the beam however, which is identified by the dashed lines in Fig.~\ref{fig:FF_Tracks}.  This section of the target edge was fit and compared to the fit of the $\alpha$-track vertices for a consistency check with the differential space charge correction.

\begin{figure}[ht]
    \centering
    \includegraphics[width=1.\linewidth]{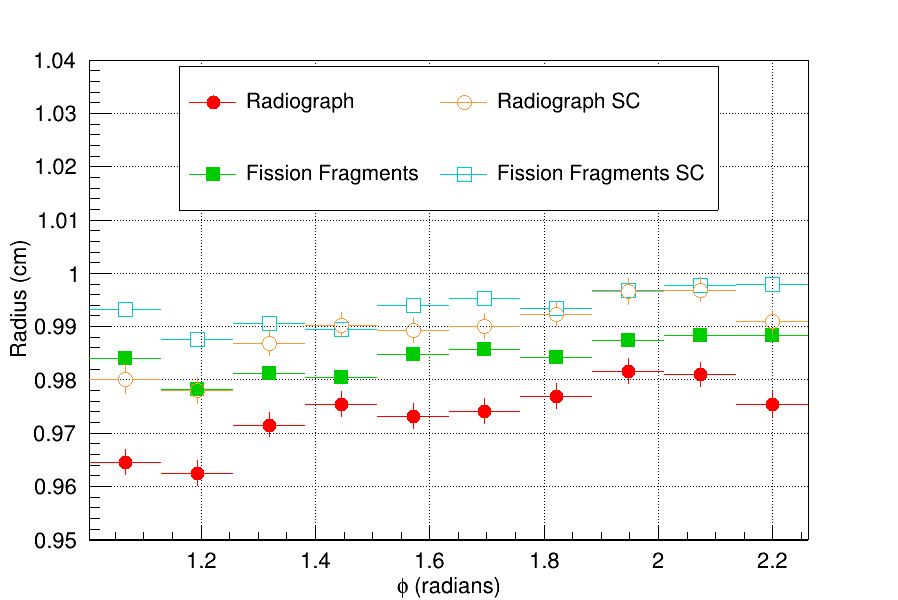}
    \caption{The target edge radius as determined by a fit with Eq.~\ref{eqn:erf} over the $\phi$ range delineated in Fig.~\ref{fig:FF_Tracks}.  The target edge is fit on $\alpha$-track radiograph and fission fragments, both with and without the space charge correction (SC) applied. The error bars are those reported by the fitting algorithm.}
    \label{fig:FF_Radiograph_Compare}
\end{figure}

Fig.~\ref{fig:FF_Radiograph_Compare} shows the results of a target edge fit vs.~$\phi$ over the range highlighted in Fig.~\ref{fig:FF_Tracks} for the $\alpha$-track radiograph and the fission fragments, both with and without the space charge correction applied.  Note that the target radius of the $\alpha$-tracks without a space charge correction applied is systematically lower than the fission fragment target edge, also without a space charge correction applied, as expected.  Also note that the application of the differential space charge from Fig.~\ref{fig:SpaceGain} brings the $\alpha$-track radiograph and fission fragment target radius into general good agreement. 

\subsubsection{Overlap Correction Summary}
While the targets used for this measurement were largely uniform, they were also of substantially different size, which necessitated an overlap correction. As part of the overlap correction a space charge effect must be taken into account, whereby the plutonium target diameter as determined by a reconstruction of $\alpha$-track vertices appears smaller than the target diameter in actuality. The space charge correction was complicated by two factors which led to additional data collection, simulation and analysis.
The first complicating factor was that the plutonium target was laid down in two separate vapor depositions, in between which a small misalignment was introduced. The misalignment resulted in the target edge being not perfectly circular, which required a more involved analysis supported by simulation to determine the space charge correction.
The second complicating factor was that the fission fragment beam data and the radiograph data were collected with different gain settings on the MICROMEGAS. This resulted in the magnitude of the space charge correction for the data subsets being different. An additional dataset was collected to quantify the change in the space charge correction as a function of gain. 

\begin{figure}[ht]
    \centering
    \includegraphics[width=1.\linewidth]{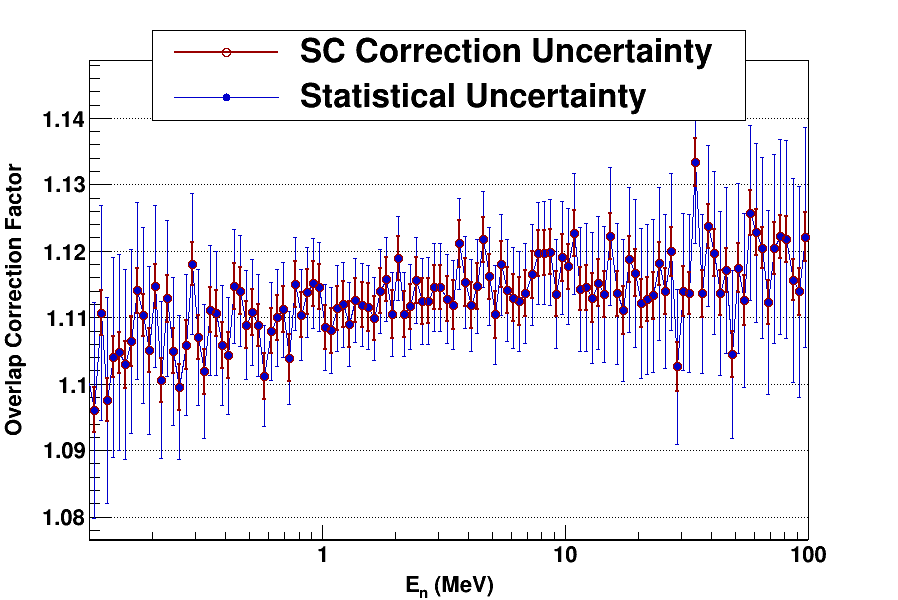}
    \caption{The overlap correction vs.~neutron energy. The error bars for statistical and space charge (SC) correction are both shown.}
    \label{fig:Overlap}
\end{figure}

The overlap correction vs.~incident neutron energy is shown in Fig.~\ref{fig:Overlap}. The overall magnitude of the correction of $\sim$11\% is a result of the plutonium target being larger than the uranium, while the trend as a function of incident neutron energy is a result of the beam shape changing with neutron energy. 
The statistical uncertainty is shown along with the uncertainty derived from the Monte Carlo variation of the slope of the differential space charge correction.  

The impact to the overlap correction from the addition of the space charge correction, and its uncertainty, is systematic with energy. This is because the shape of the beam and target remain fixed and it is just the scaling of the diameter that is changing.  The total number of counts is conserved (and is normalized to 1), therefore scaling the diameter systematically decreases each bin value. This can be thought of as decreasing the areal density of the target, or correspondingly decreasing the intensity of the beam. For the purposes of error propagation, this systematic uncertainty is added in quadrature with the uncertainty of the target mass normalization uncertainty in Sec.~\ref{sec:normalization}.

\subsection{Normalization}
\label{sec:normalization}

The absolute normalization of the previous fissionTPC measurement of the $^{239}$Pu$(n,f)/^{235}$U$(n,f)$ cross-section ratio \cite{Snyder2021jor} was determined with a combination of mass spectrometry and $\alpha$-spectroscopy measurements, detailed in Ref.~\cite{Monterial2021}. For the results described in this article the same general approach was used with one important difference: the fissionTPC radiograph data were used for the $\alpha$-spectroscopy instead of a silicon detector. Mass spectrometry results are presented in ~\ref{app:mass_spec}.

There are advantages and disadvantages when comparing a silicon detector to a gas-based detector.  A silicon detector has a lower acceptance than a 2$\pi$ gas detector, but the acceptance is well defined by the geometry of the apparatus (detector diameter, and distance from source), though detector edge effects can exist and may cause the silicon detector efficiency to be different than it is assumed based on geometry only. A 2$\pi$ ionization chamber has greater acceptance, however, its efficiency is not exactly 100\% as $\alpha$-particles can lose energy in the sample material as they approach emission angles close to parallel with the source surface. As a result, the efficiency of a 2$\pi$ ionization chamber is less well-defined as it will depend on details of the sample material makeup and detector thresholds.  Regardless of the technique used, determining the efficiency to an accuracy of $<$1\% is challenging.  The approach to this challenge in Ref.~\cite{Monterial2021} was to measure the plutonium and uranium target in a silicon detector system designed to precisely maintain the detector geometry and limit effects such as scattering.  Since the absolute normalization is a ratio of the target masses, in principle, detector efficiency is not needed as long as it is the same for both the plutonium and uranium measurements. This choice of detector setup presents its own challenge; since the half-lives of uranium and plutonium differ by $\sim$5 orders of magnitude, it is difficult to optimize the distance between source and detector to limit pileup and dead time from the plutonium while also collecting sufficient uranium statistics in a reasonable time.
 
The fissionTPC, on the other hand, is not simply an ionization chamber, but a 3D tracking detector, and using it for such an analysis has several advantages. First and foremost, it was desired that the $\alpha$-counting and beam data be collected from a single target-loading-and-assembly procedure of the fissionTPC to eliminate any chance of target handling resulting in damage and affecting the normalization measurement. 
As mentioned in Sec.~\ref{sec:introduction}, this could be a contributing factor to the discrepancy with ENDF.
Secondly, the fissionTPC effectively has no dead time and pileup can be accurately accounted for, which is detailed later in this section.  
Finally, the fissionTPC tracking allowed for a different approach to the efficiency. For the silicon detector measurement the efficiency was held fixed by the source--detector separation, and assumed to be the same for both the plutonium and uranium targets. For the fissionTPC data, the efficiency can be determined directly from the data and a selection cut can be made with an associated uncertainty. 

\subsubsection{Alpha Counting Efficiency}
Rather than attempting to quantify the absolute efficiency, the isotropic nature of $\alpha$-decay and fissionTPC particle tracking make it possible to make a cut on the polar angle in a region where 100\% of the $\alpha$-tracks escape the target material and then correct for well defined selection efficiency.  Nevertheless, the fissionTPC tracking is subject to biases that must be taken into account: the difference between lateral and transverse diffusion can bias angle determination, particularly for high gain data or high energy tracks, and field distortions, such as space charge, can also bias tracking. 

\begin{figure}[ht]
    \centering
    \includegraphics[width=1.\linewidth]{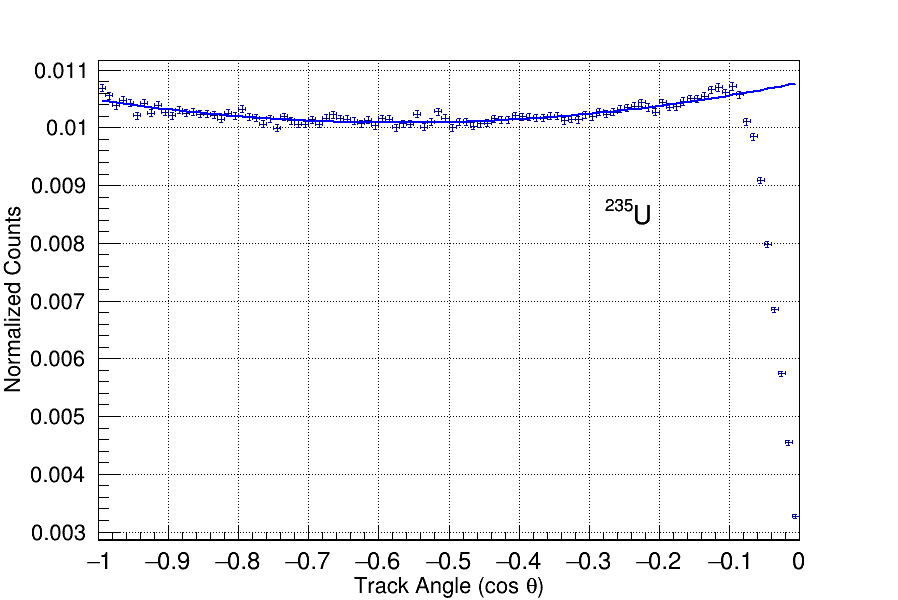}
    \includegraphics[width=1.\linewidth]{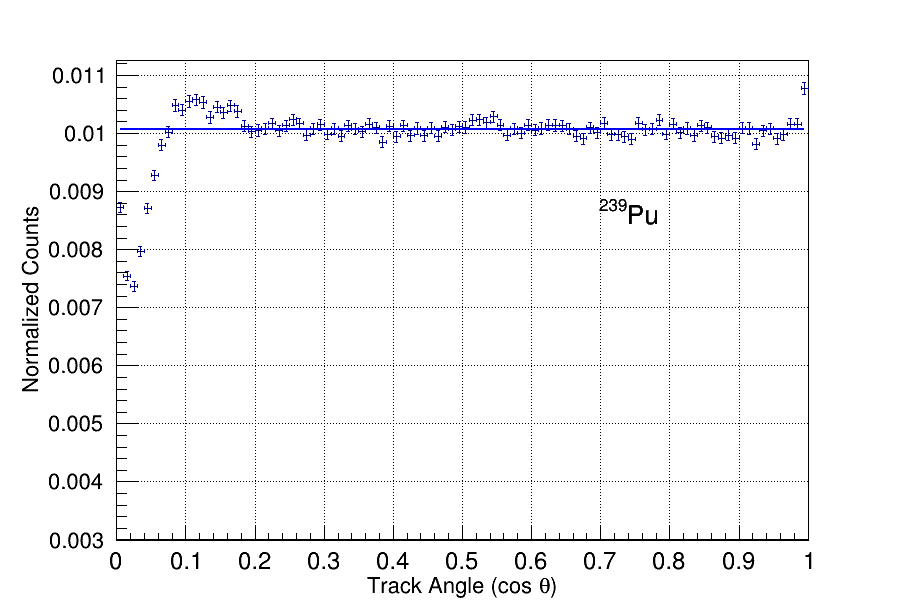}
    \caption{The polar angle distribution of $\alpha$-particle tracks from the uranium (top) and plutonium (bottom) targets.  A polynomial fit is used to make a correction for the tracking bias away from the flat distribution expected for isotropic emission.}
    \label{fig:CosTheta}
\end{figure}

Fig.~\ref{fig:CosTheta} shows the polar angle distributions for uranium and plutonium radiograph data.  Isotropic data should exhibit a flat distribution vs.~$\cos\left(\theta\right)$, where $\theta$ is the polar angle, but the uranium data exhibits a deviation from a flat distribution. 
A tracking bias in fission fragments that results from differences in lateral and transverse diffusion in the fissionTPC were previously observed~\cite{hensle2020}.  Since fission fragment tracks have a high ionization density the diffuse charge is readily observable.
The uranium data were collected at a high gain to improve pointing resolution, which also increases the amount of diffuse charge that is above threshold, and this is likely the cause of the observed tracking bias.  The bias could also be related to the field cage voltage settings discussed in Sec.~\ref{sec:gain}.  Regardless of the cause, the distortion is well characterized by a simple polynomial and can be used to correct a selection efficiency cut.  The correction to the selection efficiency is $\sim$0.5\% and is kept small due to choice of cut about the symmetric point of the distortion at $\cos\left(\theta\right) = -0.5$.  The drop in counts as they approach $\cos\left(\theta\right) = 0$ is a result of $\alpha$-particles lost in the target material which is $\sim$100 $\mu g/cm^2$ of UF$_4$, while the slight increase in counts near $\cos\left(\theta\right) = -0.1$ is the result of $\alpha$-particles back-scattering in the target and/or target backing.

The polar angle distribution for the plutonium is flatter than that of the uranium.
It is not unreasonable to expect the plutonium side to have different tracking biases than the uranium side.
The gain achievable on the plutonium side was limited to $\sim$1/3 of that of the uranium side, eliminating much of the diffuse charge as it falls below threshold. Also, the space charge effect discussed in Sec.~\ref{sec:overlap} is only present on the plutonium side.  
To be consistent, the plutonium was fit with a polynomial as well and corrected to a flat distribution, which was only a 0.04\% correction.  Note that the polar angle distribution range is opposite that of the uranium side as the total geometry of the fissionTPC is 4$\pi$ and the two deposits are on opposite sides.  The energy loss and back-scattering features on the plutonium distribution are on the left hand side near $\cos\left(\theta\right) = 0.1$.

The uncertainty of the efficiency correction was estimated with Monte Carlo realizations based on the covariance matrix provided by the fitting algorithm. The fit range was limited to $0.2<\cos\left(\theta\right)<1$ (negative on the uranium side), as the back-scattering and energy loss features near $\cos\left(\theta\right) = 0$ are not the result of a tracking bias or field distortion, and the fit is projected through this region.  A $3^{rd}$ order polynomial is used to allow for some additional variation in the fit uncertainty in the projected region, though no large deviations from a smooth distribution are anticipated. 

\subsubsection{Pileup and Dead Time}
A silicon detector system experiences pileup when two or more particles deposit energy in the detector within the shaping time of the amplifier. In the case of two-particle pileup, the resulting pulse will be measured to have an energy ranging up to twice the expected energy of a single particle (assuming a monoenergetic source), depending on how close in time the particles arrive, the shaping time of the amplifier, and choices made in how a digitized waveform is processed.  In the fissionTPC, pileup occurs when two or more particle tracks deposit ionization in the drift volume not only coinciding in time but also along a trajectory such that the track finding algorithm cannot resolve two separate tracks.  In the case of two-track pileup, the resulting composite track will have up to twice the energy. Often in these pileup scenarios, some smaller anomalous tracks will also be found, as the two true tracks rarely overlap perfectly in space. These smaller tracks are typically portions of the ends of the true tracks. Similarly, the composite track will have up to twice the length of an expected track, while anomalous tracks will be very short.  The distinct characteristics of these pileup events mean they are present in a well-separated portion of the phase space of the fissionTPC data and can be accurately accounted for.

The fissionTPC was designed to have a 100\% live time and all measurements indicate that this was achieved. A detailed description of the fissionTPC electronics can be found in Ref.~\cite{Heffner2013} while a validation of the live time can be found in Ref.~\cite{Snyder2021jor}.

\subsubsection{Uranium Spectroscopy} 
\label{sec:uranium_spec}
The energy resolution of the fissionTPC is poor compared to silicon detectors.  The track length resolution of the fissionTPC however, is substantially better than the energy resolution, and since $\alpha$-particle track lengths in a gas are proportional to their energy, they can be used for $\alpha$-spectroscopy.  While the fissionTPC track length resolution is still less than the equivalent energy resolution of a silicon detector, it is adequate for the normalization measurement presented here.

\begin{figure}[ht]
    \centering
    \includegraphics[width=1.\linewidth]{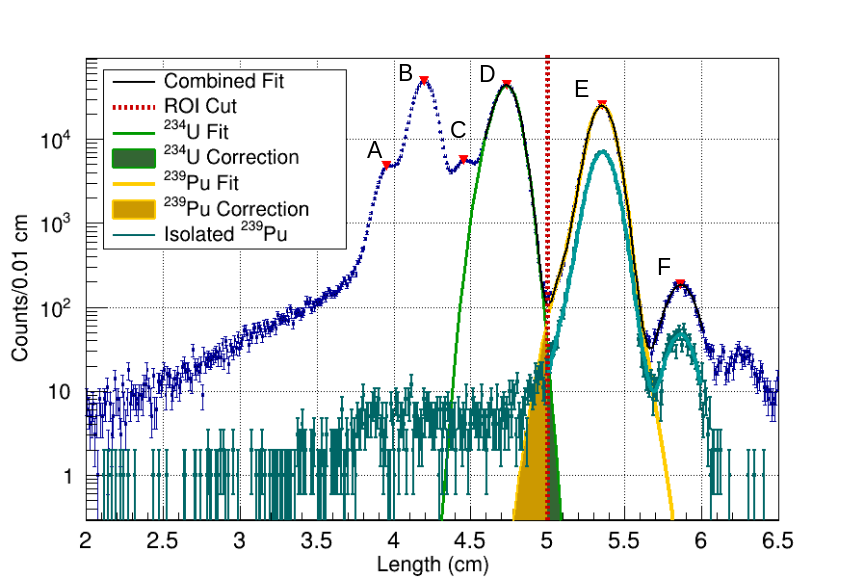}
    \caption{The $\alpha$-particle track length spectrum of the uranium target. Peaks \emph{A,B, \& C} are from $^{235}$U, peak \emph{D} is $^{234}$U, peak \emph{E} is $^{239}$Pu, and peak \emph{F} is $^{238}$Pu \& $^{241}$Am. Note that there are unresolved peaks that contribute to the fit parameters. The red, dashed vertical line represents the ROI cut, while the shaded regions are corrections to the ROI integral based on the fits. The \emph{Isolated $^{239}$Pu} data are from tracks originating on the target backing outside the uranium deposit, which are used to estimate a correction for scattered plutonium contribution to the uranium ROI.  
    What appears to be a peak at energies greater than peak \emph{F} is the result of various uranium decay chains \cite{Monterial2021}.}
    \label{fig:Uranium_Spectrum}
\end{figure}

Fig.~\ref{fig:Uranium_Spectrum} shows the $\alpha$-particle track length spectrum from the uranium radiograph data.  The spectrum is dominated by peaks from $^{235}$U and $^{234}$U, and also present is a background from $^{239}$Pu. The plutonium background is not a contamination within the uranium target material, but rather a thin layer over the target and on the aluminum target backing. This occurred during the vapor deposition process of the plutonium deposit on the other side of the target backing.  The assertion that the plutonium is a separate deposition rather than a contaminant in the uranium is supported by the fact that plutonium track vertices are detected over the entire 2 cm diameter target backing, not just within the uranium deposit area. Furthermore, the plutonium tracks do not experience the same energy loss at high $\cos\left(\theta\right)$ in comparison to the uranium tracks, suggesting they are not traveling through the $\sim$100 $\mu g/cm^2$ of UF$_4$ but rather are on top of it.  While this background will affect the normalization measurement, it has a negligible impact on the beam fission counting, as $^{239}$Pu has an $\alpha$-decay half-life 5 orders of magnitude shorter than that of $^{235}$U, indicating that the amount of uranium deposited on this side of the target is more than 5 orders of magnitude greater than the amount of plutonium, simply based on a visual assessment of the relative area under the $\alpha$-peaks. 

In general, the analysis procedure employed here first made use of a polar angle cut to remove the portion of the tracks near parallel to the target surface that are highly degraded in energy, which necessitated an efficiency correction, as already described. Then a region of interest (ROI) was defined around the uranium counts, cutting the plutonium counts which are at a longer track length (greater energy). The high energy uranium peaks and the plutonium peaks were then fit with Gaussians to correct for the number of counts that were either lost above the ROI cut or represent a background below the ROI cut.  The mass spectrometry data and DDEP \cite{DDEP} values for half-lives were then used to determine how many of the counts in the uranium ROI are from $^{235}$U. 

\begin{figure}[ht]
    \centering
    \includegraphics[width=1.\linewidth]{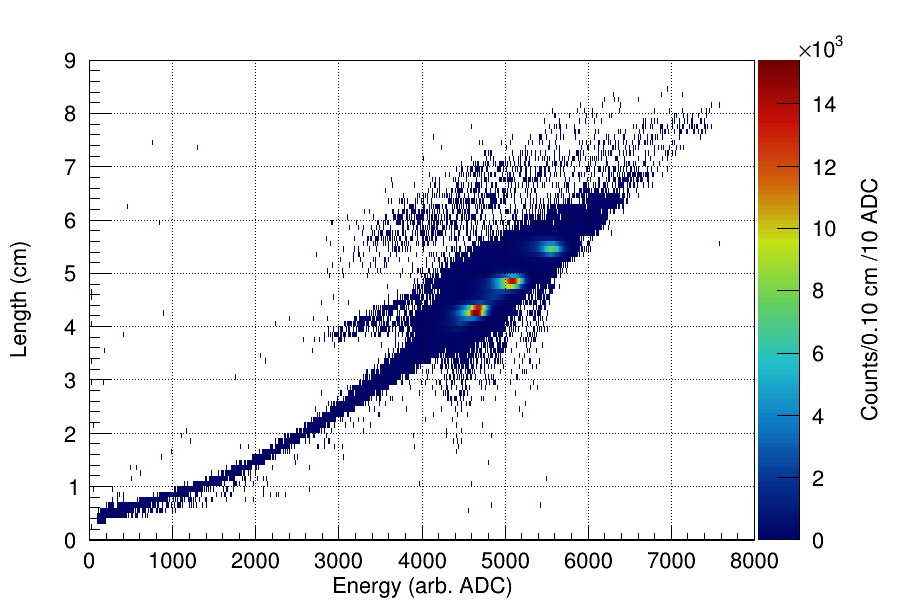}
    \caption{The $\alpha$-particle track length vs.~ energy spectrum. The tail extending to low energy is the result of $\alpha$-particles that have lost energy in hard nuclear scattering. The small population of tracks above the main $\alpha$-peaks 
    are the result of high energy tracks from the various uranium decay chains that impact the outer edge of the drift volume.}
    \label{fig:Uranium_LvA}
\end{figure}

A notable feature in Fig.~\ref{fig:Uranium_Spectrum} is the exponential-like tail that is visible at lower track lengths.  These are real tracks in the sense that they are not artifacts of the tracking algorithm, but rather $\alpha$-particles which have undergone hard nuclear scattering. The hard nuclear scatters are apparent in the plot of track length vs.~energy in Fig.~\ref{fig:Uranium_LvA}, where the low energy tail follows the expected trend of an $\alpha$-band.  These tracks are therefore included in the count. Since the tracks that undergo hard nuclear scattering are subject to large changes in angle, the number of tracks that scattered into or out of the angular cut must be accounted for. Given that the cut was placed at $\cos\left(\theta\right)$ = 0.5 or roughly a 50\% efficiency correction, the impact of the hard nuclear scattering is estimated to be negligible assuming the angular distribution of hard nuclear scatter events is isotropic.  Such an aggressive angular cut was selected because back-scattering from the target backing is the result of cumulative small angle scattering and is not isotropic.  SRIM \cite{ZIEGLER20101818} simulations indicate that a cut at $\cos\left(\theta\right)$ = 0.5 provides a more than comfortable margin from any impact of back-scattering.   

The potential for a similar exponential tail from the plutonium peak, contributing to a background in the uranium counts, that would not have been captured by a Gaussian fit, must also be taken into account.  As discussed previously, the plutonium deposit is many orders of magnitude thinner than the uranium deposit so one would not expect as substantial a scattering tail. The behavior of the plutonium in isolation can be observed by cutting on tracks with vertices outside the uranium target deposit, where only plutonium contamination is present and the tail is not obscured by uranium peaks, as shown in Fig.~\ref{fig:Uranium_Spectrum} and denoted as \emph{Isolated $^{239}$Pu}.  The correction of the scattered plutonium background is estimated by integrating the region below the ROI of the isolated plutonium spectrum and scaling it by the ratio of the areas under the $^{239}$Pu peak in the isolated data and the $^{239}$Pu peak within the uranium target data. The scaling factor is approximately 3.5. 

Tab.~\ref{table:Norm_Uncert} lists the uncertainty contributions.
The uncertainty for the corrections that are based on fits were estimated by generating Monte Carlo realizations of the fit parameters using the covariance matrices provided by the fitting algorithm. All fits were done within the ROOT analysis framework~\cite{ROOT}.

The fissionTPC has a 50 MHz clock, with every data-run length accounted for and every track within a run assigned a time. The total activity on the uranium side of the fissionTPC is $\sim$10 Bq and pileup is negligible. 

\subsubsection{Plutonium Spectroscopy}
\label{sec:plutonium_spec}
\begin{figure}[ht]
    \centering
    \includegraphics[width=1.\linewidth]{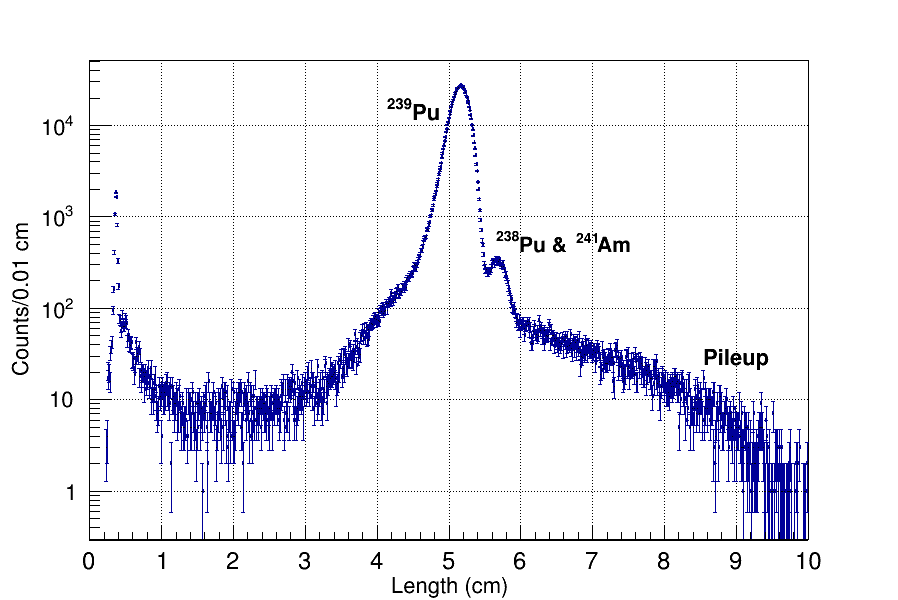}
    \caption{The $\alpha$-particle track length spectrum for the plutonium target. The large and small peaks are $^{239}$Pu and $^{238}$Pu respectively. The distribution of tracks above the peaks is the result of pileup. }
    \label{fig:Plutonium_Spectrum}
\end{figure}

The main components of the plutonium spectrum, seen in Fig.~\ref{fig:Plutonium_Spectrum}, are peaks from $^{239}$Pu and $^{238}$Pu, while this particular plutonium material also has an $^{241}$Am contribution unresolvable with the $^{238}$Pu.

\begin{figure}[ht]
    \centering
    \includegraphics[width=0.95\linewidth]{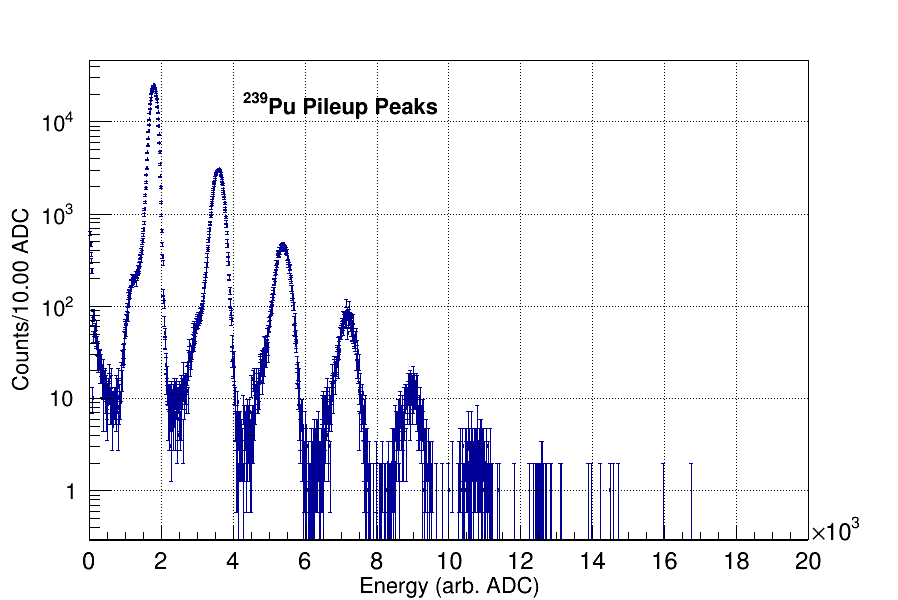}
    \caption{The total combined energy of pileup events. The distinct populations or ``peaks'' are at integer multiples of the main $\alpha$-peak energy, indicating double, triple, etc. pileup events.}
    \label{fig:Plutonium_EventADC}
\end{figure}

The overall activity of the plutonium source was $\sim$125 MBq, causing a substantial pileup rate. As discussed previously in this section, the pileup events have distinct characteristics which allow them to be accurately accounted for.
The tracks extending beyond $\sim$6 cm in Fig.~\ref{fig:Plutonium_Spectrum} are a result of the pileup events. 
As pileup tracks rarely perfectly align in the drift volume of the fissionTPC, the tracking algorithm still typically finds multiple tracks, with one representing the main body of the charge cloud and the rest being smaller portions of the cloud where the pileup tracks diverge.  These smaller tracks populate the region of Fig.~\ref{fig:Plutonium_Spectrum} below 1 cm. Fig.~\ref{fig:Plutonium_EventADC} shows a plot of the total energy deposited in all events with any multi-track event identified having all of the energy summed into a single count. A distinct pattern is visible, with well-separated populations at integer multiples of the plutonium $\alpha$-peak energy, indicating double, triple, etc. pileup events.  These populations can be selected and a true track count can be determined regardless of the number of tracks identified by the algorithm and without relying on the length distribution in Fig.~\ref{fig:Plutonium_Spectrum}.  The pileup events represent $\sim$1.5\% of the total number of track counts. The uncertainty  of the pileup correction is driven by the ambiguity of associating counts-to-pileup in the regions between the peaks. The uncertainty of the pileup correction was estimated to be 4\% of the total pileup correction, which is a $\sim$0.1\% uncertainty contribution to the final result, and was determined by varying the position of the cut selection between pileup peaks.

\begin{figure}[ht]
    \centering
    \includegraphics[width=1.\linewidth]{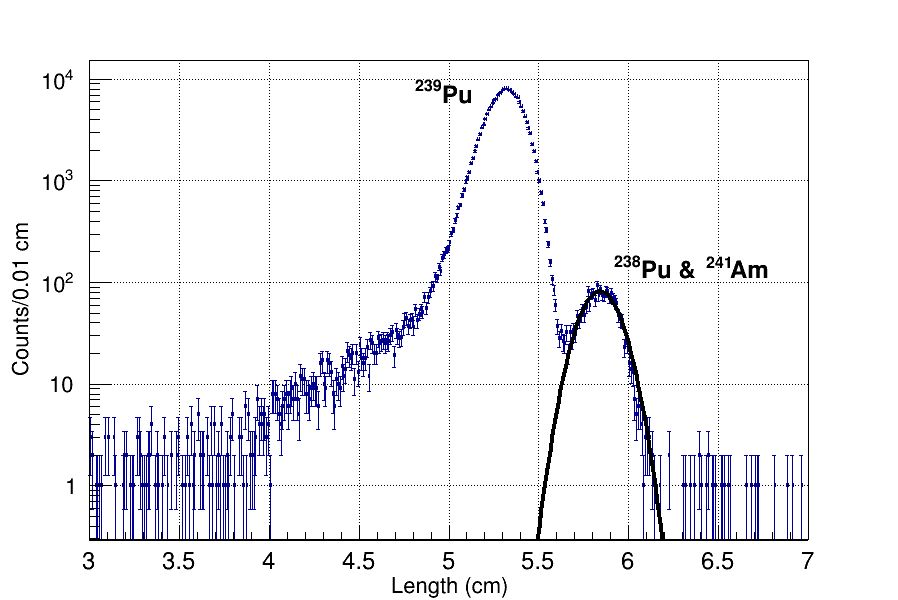}
    \caption{The plutonium $\alpha$-spectrum, with pileup events removed. The $^{238}$Pu and $^{241}$Am peak is fit to determine its relative contribution to the total decay rate.}
    \label{fig:Plutonium_AmFit}
\end{figure}

The $^{238}$Pu and $^{241}$Am were not included in the mass spectrometry measurement and therefore must be taken into account with the $\alpha$-spectroscopy measurement. Plotting only single-track events results in a cleaner length spectrum as seen in Fig.~\ref{fig:Plutonium_AmFit}. The higher energy peak is from the $^{238}$Pu and $^{241}$Am contaminants, which are unresolvable at the fissionTPC resolution. Fitting the contamination peak and integrating the ROIs determines the relative ratio of the contamination to the $^{239}$Pu, assuming that the pileup event rate that was cut is the same for both peaks.   
The overall contribution of the contaminants to the plutonium spectrum is $\sim$1\%, which was also confirmed with a high resolution silicon measurement.

The pileup rate could potentially have a greater impact on the $^{241}$Am peak relative to that of the $^{239}$Pu, which would affect the overall determination of the $^{241}$Am correction.  The spectrum in Fig. 26 that was fit included single track events only (no pileup).  Assuming the pileup rate was the same for all $\alpha$-particle energies holds when considering pileup in a strictly temporal sense, but the $^{241}$Am $\alpha$-tracks are longer and therefore may have a greater chance of crossing with another track, which is the driving force for the observed pileup in the fissionTPC. The tracks from $^{241}$Am are about 10\% longer than those from $^{239}$Pu. Assuming a 10\% higher pileup rate for $^{241}$Am would translate to a $\sim$0.7\% impact on the $^{241}$Am correction, which is again within the overall estimated uncertainty of the correction and would have a negligible impact on the final result. 

The fit in Fig.~\ref{fig:Plutonium_AmFit} ignored contributions of an exponential scattering tail from the $^{241}$Am peak under the $^{239}$Pu peak.  This is a second order correction that is estimated to have an upper limit of a 2\% impact on the pileup correction, or a 0.02\% impact on the final result, which is considered negligible and is within the error estimated for the pileup correction based on the fit covariance matrix. 

\begin{table}[ht]
\caption{Uncertainty contributions for mass normalization in \% of measured number of target atoms.}
\label{table:Norm_Uncert}
\centering
\begin{tabular}{  m{0.5\columnwidth}  m{0.2\columnwidth} m{0.2\columnwidth}  }
\hline
Source of Uncertainty & $^{235}$U  & $^{239}$Pu \\ \hline
Counting statistics & 0.075\% & 0.105\% \\
Spectrum fit  & 0.006\% & 0.023\% \\
Pileup  & NA & 0.060\% \\
Selection efficiency  & 0.202\% & 0.061\% \\
Mass spectrometry  & 0.657\% & 0.005\% \\
Half lives & 0.136\% & 0.046\%\\
\hline
Total & 0.682\% & 0.145\%\\
\hline
\end{tabular}
\end{table}

The mass normalization ratio $N_{U}/N_{Pu}$ is equal to 1.0916 $\pm$ 0.00785. 
Table~\ref{table:Norm_Uncert} lists the major uncertainty contributions.  This uncertainty is larger than the 0.05\% uncertainty reported for the previous fissionTPC cross-section ratio measurement \cite{Monterial2021}, but we have greater confidence that there was no systematic error caused by target handling.
Recall also from Sec.~\ref{sec:overlap} that there is a systematic uncertainty of 0.34\% from the overlap correction that is computed in quadrature with this uncertainty, for a total normalization uncertainty of $\pm$ 0.00857, which constitutes a 0.79\% uncertainty contribution to the final result.

\section{Cross-section Ratio Comparison and Discussion}
\label{sec:compare}

\begin{figure*}[ht]
\centering
\includegraphics[width=0.75\linewidth]{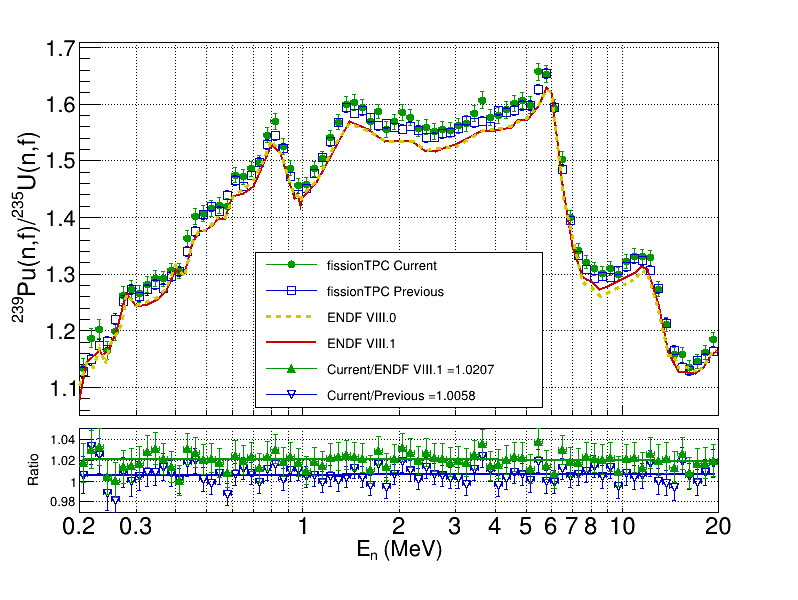}
\caption{The \ftpc{} current and previous measurements of the \pu(n,f)/\u(n,f) cross-section ratio as a function of neutron energy compared to the ENDF/B-VIII.1 evaluation. ENDF/B-VIII.0 is also shown. The lower panel shows the ratio between the current data and ENDF/B-VIII.1 evaluation and also the current relative to the previous results. The lines in the lower panel represent a $1^{st}$ order polynomial fit of the ratios. The values reported in the legend are the intercepts of those fits.  }
\label{fig:fTPC_ENDF8}
\end{figure*}

\begin{figure*}[ht]
\centering
\includegraphics[width=0.75\linewidth]{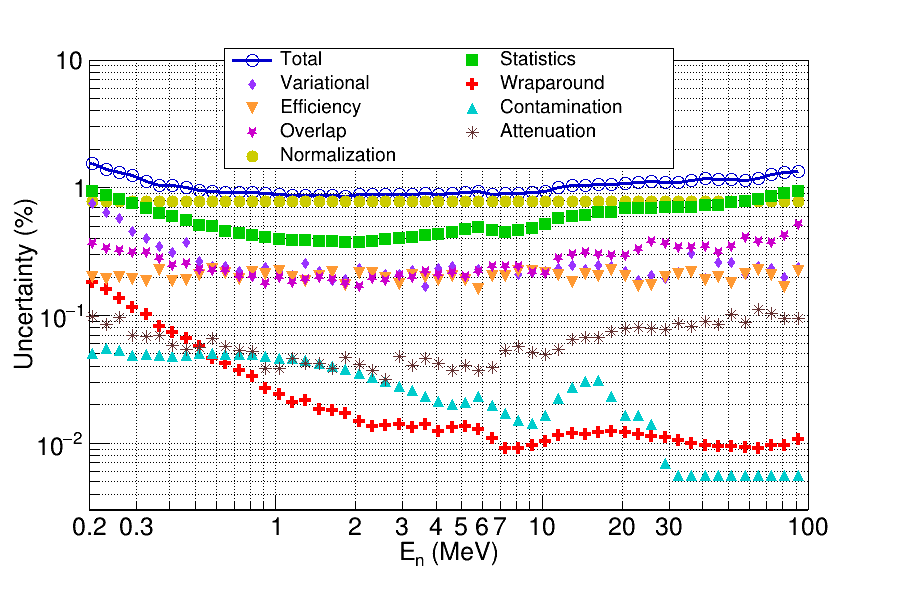}
\caption{The partial uncertainties for the current \ftpc{} measurement of the \pu(n,f)/\u(n,f) cross-section ratio as a function of neutron energy.}
\label{fig:partial_error}
\end{figure*}

Results of the current and previous fissionTPC \pu(n,f)/\u(n,f) cross-section ratio measurements as a function of neutron energy from 0.2 to 20 MeV are shown in Fig.~\ref{fig:fTPC_ENDF8}. Also shown is the ENDF/B-VIII.0 \cite{ENDF8} and ENDF/B-VIII.1 evaluations. The lower panel of the figure shows the ratio between the data and the evaluation and also the ratio between the current and previous fissionTPC results. Fig.~\ref{fig:partial_error} shows the total and partial uncertainties as a function of neutron energy.

\subsection{Discussion}

The current results are in agreement with the previous \ftpc{} results within the uncertainties.  
Although correlations between the two measurements are present, for simplicity, the error bars in the lower panel of Fig.~\ref{fig:fTPC_ENDF8} assume that the fissionTPC measurements are fully uncorrelated. Thus, these error bars represent a propagation of uncertainties for a simple ratio assuming the total uncertainties of each measurement are the partial uncertainties of a ratio.
The line in the lower panel of Fig.~\ref{fig:fTPC_ENDF8}
represents a $1^{st}$ order polynomial fit to illustrate the good shape agreement and quantify the normalization offset. The slope of the line is $2.5 \times 10^{-5}$, effectively flat, and the intercept is 1.0058 or 0.58\%.  This normalization offset is within the overall uncertainty and the reported mass normalization uncertainty for the current results of 0.79\%. 
Overall the uncertainties of the current work are larger than our previous measurement.  This is largely driven by the higher overall normalization uncertainty. A total uncertainty of less that 1\% was achieved within the energy range of 0.5 -- 10 MeV, with uncertainties near 1.1\% at 30 MeV.

The current fissionTPC results clearly support the observed systematic offset of the previous results relative to ENDF.  However, it is fair to consider what potential there is for a shared systematic error between the two measurements by examining those factors that remained the same and those that are different.  The next sections summarize these conditions.

\subsubsection{Unchanged General Detector Configurations}

Configuration changes were minimized so that effects of target uniformity along with the overlap correction, presented in Sec.~\ref{sec:overlap}, and the absolute mass normalization presented in Sec.~\ref{sec:normalization}, would be the independent variables.  The absolute mass normalization is intrinsically linked to the new uniform target however, so these two changes could not be studied in  a truly independent fashion.    

The beam line, 90L at LANSCE-WNR, was the same and the collimation was identical.  The mounting table for the fissionTPC is locked into fiducials permanently secured to the floor of 90L so that the position relative to the collimation and the beam-path length were unchanged.  The tungsten spallation target had been changed in the intervening period, so some small change in beam shape is possible, though procedures at LANSCE are in place to minimize any changes.  The 800 MeV proton accelerator itself is subject to repairs and the variable operating conditions that any accelerator must contend with.  

The design of the fissionTPC was unchanged. Some parts of the fissionTPC are interchangeable, such as the DAQ electronics and the pad plane with gain stage, as discussed in Sec.~\ref{sec:gain}, but generally the parts are intended to be identical. The gas flow system hardware and settings were also unchanged.

The analysis framework is maintained in an version controlled repository and no changes of note were recorded.  In particular, the methods employed for particle tracking, time-of-flight determination, the efficiency correction, and the wrap-around correction were all unchanged. There was some rollover in collaboration team members, but senior personnel in leadership and those primarily responsible for the analysis framework remain in their positions on the team.

The two measurements are, for all the above reasons, correlated in many ways and could be subject to shared systematic error.  It was the systematic nature (as a function of neutron energy) of the deviation with ENDF however, that pointed to the overlap correction or absolute normalization as being the potential cause, since the efficiency and wraparound corrections both have a strong dependence with neutron energy.  

\subsubsection{Changed Experimental Configurations}

Whereas the beamline and detector configuration were unchanged, a deliberate decision was made to have two uniform vapor deposited targets. This difference was made so that the $\sim2\%$ discrepancy could be addressed with uniform targets, which was not the case in the previous measurement. As a result new corrections to the overlap, space charge and normalization had to be made.

The overlap correction has an energy dependence (see Fig.~\ref{fig:Overlap}) due to the fact that the beam shape changes with energy, but it also has a normalization component, as the calculation is normalized to the area of the targets. In particular, the space charge correction (detailed in Sec.~\ref{sec:gain}), which changes the observed area of the target, but not the observed $\alpha$ or fragment counts, would in effect increase the observed intensity of the beam per unit area and shift the overlap correction systematically with energy if it were not taken into account.  The need for a space charge correction is a shared aspect of the current and previous fissionTPC measurements. In the work presented here however, a greater detail of analysis was required due to the change in gain settings, but also a supporting and validating dataset for the space charge correction was collected. 

With a new target came the need for a new normalization measurement. The method employed for this measurement and the previous work shared the same general approach; count $\alpha$-decays and correct with a mass spectrometry measurement in combination with published half-lives. 
The techniques and facilities for the mass spectrometry were the same for both measurements, as were the half-lives used in the analysis, which could contain a shared systematic error. The source material for the targets, however, was not the same. Specifically, the uranium used in the previous measurement contained $^{233}$U which accounted for about 1/3 of the total activity.  There was no detectable $^{233}$U in the current data. The current plutonium target also contained $^{241}$Am, which resulted from the beta decay of $^{241}$Pu, while no $^{241}$Pu was detected in the previous target.  The detailed history of these source materials is not known.   
The technique employed for $\alpha$-counting was different for the two measurements.  A silicon detector was used in the previous work, while the fissionTPC itself was used for the current measurement (see Sec.~\ref{sec:normalization}). 

\subsection{Conclusions}
The results of the fissionTPC measurement of the $^{239}$Pu(n,f)/$^{235}$U(n,f) cross-section ratio published in 2021 showed a deviation from the ENDF/B-VIII.1 evaluation, systematic as a function of incident neutron energy of just under 2\% \cite{Snyder2021jor}. Some experimental challenges, namely a highly nonuniform plutonium target in conjunction with a nonuniform beam, led the NIFFTE collaboration to the conclusion that the cross-section ratio should be remeasured with the fissionTPC to address the potential for systematic errors to be the cause of the discrepancy.

A uniform target was produced, another measurement was made, and the results have been presented herein.  The current results are in agreement with the previous measurement within uncertainties. The shape agreement is excellent, with the central values being, in fact,  slightly more systematically discrepant with ENDF. This new measurement faced its own unique challenges, and as such, the analysis and quantification of uncertainties has been as exhaustive as possible, as was the case in our previous measurement.

It is our conclusion that these results should be treated by evaluators as being absolutely normalized.  We also conclude that with the substantive differences between our current and previous experiments leading to the same result, these measurements should be treated as cross validating one another, and our previous results should be considered absolutely normalized as well. It would be conservative to treat them as fully correlated measurements.

While it is perhaps obvious to those with experience in this field, it is worth emphasizing that the careful production and metrology of targets for this type of experiment is of the utmost importance. It is interesting to note that, generally speaking, publications for fission cross-section ratios focus on fission chambers and neutron beams, while target normalization is often only cursorily addressed.
While much work has already been done on reviewing previous $^{239}$Pu(n,f) measurements and their uncertainties \cite{Neudecker2020}, it may be worth investigating the potential of systematic errors in previous measurements specifically related to normalization.  

\section{Acknowledgments}
The authors would like the thank A.M.~Gaffney (LLNL) for providing mass spectrometry results.
University collaborators acknowledge support for this work from the DOE-NNSA Stewardship Science Academic Alliances Program, under Grant No.~DE-NA0002921, and through sub-contracts from LLNL and LANL. 
The neutron beam for this work was provided by LANSCE, which is funded by the U.S. Department of Energy and operated by Los Alamos National Security, LLC, under Contract No.~89233218CNA000001.
This work performed under the auspices of the U.S. Department of Energy by Lawrence Livermore National Laboratory under Contract No.~DE-AC52-07NA27344. 
Cleared for external release, distribution unlimited LLNL-JRNL-2000446.


\twocolumn
\bibliographystyle{elsarticle-num}
\bibliography{references.bib}









\newpage
\appendix

\section{Fission Fragment Selection Cuts}
\label{app:frag_selection}

\begin{figure}[ht]
 \centering
 \includegraphics[width= 1.\linewidth]{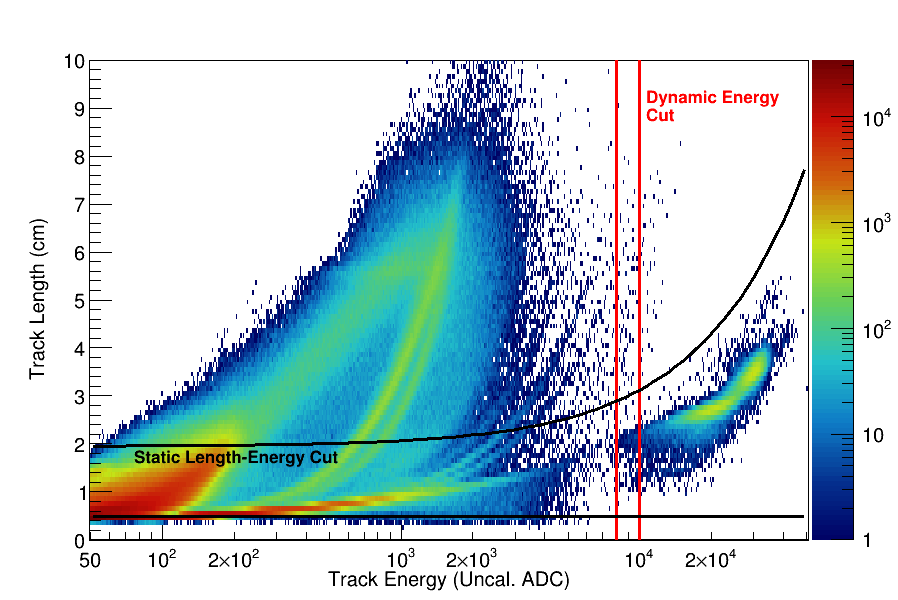}
 \caption{Selection cuts applied to the length vs.~track energy distribution of selected particles. Two static cuts remove background contributions, while one dynamic cut is varied within the range of 7500 -- 10000 ADC. The dynamic cut is used to determine residual uncertainties from the efficiency correction.}
 \label{fig:lvadc_labelled}
\end{figure}

Excellent particle identification (PID) is a central capability of the \ftpc{}, which derives
primarily from track reconstruction. Track length (L) plotted against track energy (E) (Fig.~\ref{fig:lvadc_labelled}) is the primary parameter space used for PID. In order to select fission fragments the following cuts, beyond a simple minimum energy threshold, are applied~\cite{Snyder2021jor}:

\begin{itemize}
 \item require track start vertices to reside within the fiducial target area,
 \item require reconstructed track length $\geq0.5$ cm in order to remove the low-energy tail,
 \item require an energy-dependent reconstructed track length maximum, which removes overlapping fragment tracks above the main distribution,
 \item require a reconstructed track polar angle of cos($\theta$) $\geq$ 0.2 to remove tracks that have the greatest amount of straggling in the target material, 
 \item accept only events that have a valid neutron time-of-flight,
 \item finally, apply cuts requiring that a given fragment track has a corresponding incident-neutron pulse.
\end{itemize}

The requirement that a fragment be associated with a valid neutron time-of-flight is important so that the incident-neutron energy can be determined. Furthermore, this requirement suppresses contamination that arises from the background contributions.

Visual representations of the fission fragment selection cuts applied to the $L$~vs.~$E$ parameter space are displayed in Fig.~\ref{fig:lvadc_labelled}. Two static PID cuts remove non-fragment backgrounds, while a dynamic cut is used to estimate residual uncertainties in the fission fragment selection efficiency.
The dynamic cut procedure is part of the variational analysis used to estimate residual uncertainties as described in \cite{Snyder2021jor}.

\begin{figure}[ht]
\centering
\includegraphics[width= 1.\linewidth]{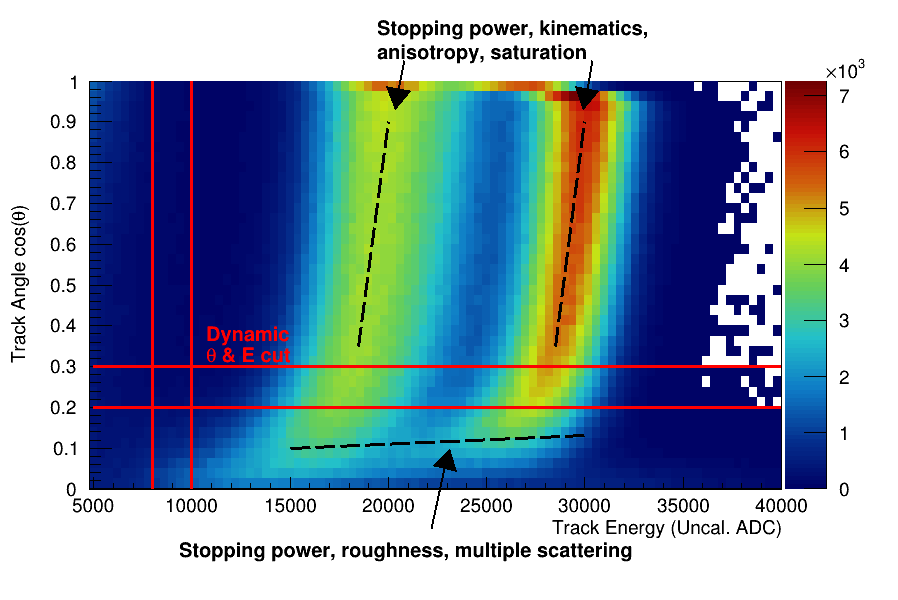}
\caption{\label{fig:Angle_vs_Energy_Explain} Fission fragment energy as a function of angle with respect to the target plane, where $\cos(\theta)=1$ is perpendicular to the plane. The two bands represent the light and heavy fission fragments. Labels are included to show the physical effects that control the shape of the fragment bands, in particular how they vary as a function of $\cos(\theta)$. A selection cut is applied to lower $\cos(\theta)$ and ADC values as the myriad effects that control the shape in this region cannot be accurately modeled at this time. The dynamic cuts are used in a variational analysis to estimate residual uncertainties and validate the method.}
\end{figure}

The fission fragment detection efficiency changes as fragment energy cuts are applied, principally due to variable energy loss in the target as a function of the track emission angle with respect to the target surface, $\cos(\theta)$. A fragment track traveling perpendicular to the target plane, $\cos(\theta)=1$, will have minimal energy loss, while one traveling nearly parallel to the plane can have significant enough energy loss as to prevent it from being detected, by either falling below the minimum energy cut or stopping completely in the target. The energy angle relationship can be observed in Fig.~\ref{fig:Angle_vs_Energy_Explain}, where the distribution bends towards lower energies as the angle approaches a value of $\cos(\theta)=0$.  The angle vs.~energy distribution is the primary feature of the \ftpc{} data that is used to constrain the efficiency model.  

At forward emission angles which are nearly perpendicular to the target, as well as emission angles approximately parallel with the target, the primary processes that affect the shape of these distributions have been described in~\cite{Casperson2018,Snyder2021jor}.

For this \pu{}/\u{} dataset the dynamical threshold cuts were employed on the angle and energy of the fragments with lower bounds of $\cos(\theta)\geq$ 0.2 and ADC $\geq$ 8500. Essentially the ``clean'' portion of the distribution at forward angles was used to determine the effect of kinematic boost and fission anisotropy and the efficiency model was used to calculate the fraction of events lost at low  $\cos(\theta)$ values. This means that the kinematic anisotropy effects are the primary determinant for how fission fragments are distributed in $\cos(\theta)$. By obtaining agreement between the model and data in a broad $\cos(\theta)$ region where straggling and target surface effects are sub-dominant, accounting for the effect of a selection cut excluding upper and lower bounds on $\cos(\theta)$ can be conducted correctly. 
\onecolumn
\section{Mass Spectrometry Table}
\label{app:mass_spec}
Tablulated results of the isotopic percentages of the uranium and plutonium targets are shown in Tab.~\ref{table:Isotopics}. 
Mass spectrometry of surrogate targets was used to determine the isotopic concentrations in the uranium and plutonium targets using the same facilities and techniques described in our previous work \cite{Monterial2021}.  The surrogate targets were prepared at OSU in the same manner and time frame as the targets used for the beam measurement.
Three samples were prepared from a single source by pipetting a small volume of acid onto the source and removing it to a separate vial.
Sample solutions were purified using standard methods and analyzed by MC-ICP-MS.
Expanded uncertainties are two times standard uncertainties, and represent an approximate 95\% confidence interval.
There is a slight variation in the $^{238}$U/$^{235}$U (3x the reported uncertainty) ratio that is consistent with trace variation in the natural U background on the source or in the mass spectrometer.
$^{233}$U was not detected and $^{238}$Pu cannot be accurately measured in the equipment used due to $^{238}$U contamination.
\begin{table}[ht]
\begin{center}
\caption{Uranium and plutonium target isotopic atom percentages as measured with mass spectrometry. }
\label{table:Isotopics}
\begin{tabular}{ccccccc}
\hline
Isotope    & Sample \#1 & Uncertainty & Sample \#2 & Uncertainty & Sample \#3 & Uncertainty \\ \hline
$^{234}$U  & 0.02891    & 0.00042     & 0.02894    & 0.00042     & 0.02895    & 0.00041     \\
$^{235}$U  & 99.89375   & 0.00081     & 99.89159   & 0.00075     & 99.89276   & 0.00072     \\
$^{236}$U  & 0.01586    & 0.00048     & 0.01581    & 0.00048     & 0.01586    & 0.00047     \\
$^{238}$U  & 0.06149    & 0.00049     & 0.06366    & 0.00037     & 0.06243    & 0.00037     \\ \hline
Isotope    & Sample \#1 & Uncertainty & Sample \#2 & Uncertainty & Sample \#3 & Uncertainty \\ \hline
$^{239}$Pu & 0.999356   & 0.000013    & 0.9993324  & 0.0000095   & 0.9993419  & 0.0000094   \\
$^{240}$Pu & 0.0004721  & 0.0000078   & 0.0004816  & 0.0000071   & 0.0004773  & 0.0000069   \\
$^{241}$Pu & 0.0000708  & 0.0000034   & 0.0000725  & 0.0000033   & 0.0000705  & 0.0000031   \\
$^{242}$Pu & 0.0001007  & 0.0000099   & 0.0001135  & 0.0000055   & 0.0001104  & 0.0000055  \\
\hline
\end{tabular}
\end{center}
\end{table}

\section{Cross-section Ratio Results Table}
\label{app:Results}
Tabulated fission cross-sections of \pu{} relative to \u{} over a neutron energy range of 0.1 -- 100~MeV are given here.  The bin structure is logarithmic.  The reported energy bin value is the lower edge of the bin.  The absolute, total uncertainty is 1~$\sigma$. Results are discussed in Sec.~\ref{sec:compare}.

\begin{longtable}{c|c|c|c|c}
\caption{The $^{239}$Pu(n,f)/$^{235}$U(n,f) cross-section ratio results.}
\label{table:results}\\
\hline\hline
Energy Bin (MeV) & Bin Width & $\sigma_{(n,f)}$ (ratio) & Uncertainty (abs.) & Uncertainty (\%) \\  
\hline
\endfirsthead
\hline\hline
Energy Bin (MeV) & Bin Width & $\sigma_{(n,f)}$ (ratio) & Uncertainty (abs.) & Uncertainty (\%) \\
\hline
\endhead
\hline\hline
\endfoot
0.1059 & 0.00628 & 0.963 & 0.028 & 2.9 \\
0.1122 & 0.00665 & 0.95 & 0.026 & 2.7 \\
0.1189 & 0.00704 & 0.9957 & 0.024 & 2.5 \\
0.1259 & 0.00746 & 0.9815 & 0.023 & 2.4 \\
0.1334 & 0.0079 & 1.045 & 0.021 & 2.1 \\
0.1413 & 0.00837 & 1.071 & 0.023 & 2.2 \\
0.1496 & 0.00887 & 1.098 & 0.021 & 1.9 \\
0.1585 & 0.00939 & 1.083 & 0.02 & 1.8 \\
0.1679 & 0.00995 & 1.091 & 0.019 & 1.8 \\
0.1778 & 0.0105 & 1.13 & 0.018 & 1.6 \\
0.1884 & 0.0112 & 1.079 & 0.018 & 1.7 \\
0.1995 & 0.0118 & 1.134 & 0.017 & 1.5 \\
0.2113 & 0.0125 & 1.187 & 0.018 & 1.5 \\
0.2239 & 0.0133 & 1.202 & 0.017 & 1.4 \\
0.2371 & 0.0141 & 1.166 & 0.016 & 1.4 \\
0.2512 & 0.0149 & 1.198 & 0.016 & 1.3 \\
0.2661 & 0.0158 & 1.263 & 0.015 & 1.2 \\
0.2818 & 0.0167 & 1.272 & 0.016 & 1.2 \\
0.2985 & 0.0177 & 1.266 & 0.015 & 1.2 \\
0.3162 & 0.0187 & 1.282 & 0.014 & 1.1 \\
0.3350 & 0.0198 & 1.293 & 0.014 & 1.1 \\
0.3548 & 0.021 & 1.292 & 0.013 & 1 \\
0.3758 & 0.0223 & 1.306 & 0.013 & 1 \\
0.3981 & 0.0236 & 1.305 & 0.014 & 1 \\
0.4217 & 0.025 & 1.363 & 0.014 & 1 \\
0.4467 & 0.0265 & 1.402 & 0.014 & 1 \\
0.4732 & 0.028 & 1.407 & 0.014 & 0.98 \\
0.5012 & 0.0297 & 1.416 & 0.013 & 0.95 \\
0.5309 & 0.0315 & 1.422 & 0.014 & 0.97 \\
0.5623 & 0.0333 & 1.42 & 0.013 & 0.93 \\
0.5957 & 0.0353 & 1.475 & 0.013 & 0.91 \\
0.6310 & 0.0374 & 1.472 & 0.013 & 0.91 \\
0.6683 & 0.0396 & 1.486 & 0.014 & 0.92 \\
0.7079 & 0.0419 & 1.496 & 0.014 & 0.92 \\
0.7499 & 0.0444 & 1.546 & 0.014 & 0.9 \\
0.7943 & 0.0471 & 1.569 & 0.014 & 0.92 \\
0.8414 & 0.0499 & 1.525 & 0.014 & 0.91 \\
0.8913 & 0.0528 & 1.487 & 0.013 & 0.91 \\
0.9441 & 0.0559 & 1.456 & 0.013 & 0.86 \\
1.000 & 0.0593 & 1.458 & 0.013 & 0.89 \\
1.059 & 0.0628 & 1.486 & 0.013 & 0.89 \\
1.122 & 0.0665 & 1.504 & 0.013 & 0.86 \\
1.189 & 0.0704 & 1.54 & 0.013 & 0.87 \\
1.259 & 0.0746 & 1.568 & 0.014 & 0.86 \\
1.334 & 0.079 & 1.6 & 0.014 & 0.89 \\
1.413 & 0.0837 & 1.603 & 0.014 & 0.88 \\
1.496 & 0.0887 & 1.594 & 0.014 & 0.86 \\
1.585 & 0.0939 & 1.569 & 0.014 & 0.88 \\
1.679 & 0.0995 & 1.587 & 0.013 & 0.83 \\
1.778 & 0.105 & 1.555 & 0.013 & 0.86 \\
1.884 & 0.112 & 1.569 & 0.013 & 0.86 \\
1.995 & 0.118 & 1.585 & 0.014 & 0.88 \\
2.113 & 0.125 & 1.576 & 0.014 & 0.87 \\
2.239 & 0.133 & 1.557 & 0.014 & 0.88 \\
2.371 & 0.141 & 1.56 & 0.014 & 0.87 \\
2.512 & 0.149 & 1.552 & 0.014 & 0.89 \\
2.661 & 0.158 & 1.555 & 0.014 & 0.88 \\
2.818 & 0.167 & 1.554 & 0.014 & 0.88 \\
2.985 & 0.177 & 1.562 & 0.013 & 0.86 \\
3.162 & 0.187 & 1.565 & 0.014 & 0.89 \\
3.350 & 0.198 & 1.584 & 0.014 & 0.9 \\
3.548 & 0.21 & 1.607 & 0.015 & 0.91 \\
3.758 & 0.223 & 1.576 & 0.014 & 0.89 \\
3.981 & 0.236 & 1.581 & 0.014 & 0.89 \\
4.217 & 0.25 & 1.59 & 0.014 & 0.91 \\
4.467 & 0.265 & 1.601 & 0.015 & 0.91 \\
4.732 & 0.28 & 1.607 & 0.015 & 0.9 \\
5.012 & 0.297 & 1.598 & 0.015 & 0.92 \\
5.309 & 0.315 & 1.657 & 0.015 & 0.92 \\
5.623 & 0.333 & 1.653 & 0.016 & 0.94 \\
5.957 & 0.353 & 1.592 & 0.015 & 0.95 \\
6.310 & 0.374 & 1.503 & 0.013 & 0.88 \\
6.683 & 0.396 & 1.4 & 0.013 & 0.92 \\
7.079 & 0.419 & 1.341 & 0.012 & 0.9 \\
7.499 & 0.444 & 1.321 & 0.012 & 0.88 \\
7.943 & 0.471 & 1.309 & 0.012 & 0.91 \\
8.414 & 0.499 & 1.302 & 0.012 & 0.95 \\
8.913 & 0.528 & 1.31 & 0.012 & 0.92 \\
9.441 & 0.559 & 1.3 & 0.013 & 0.97 \\
10.00 & 0.593 & 1.322 & 0.012 & 0.94 \\
10.59 & 0.628 & 1.332 & 0.013 & 0.97 \\
11.22 & 0.665 & 1.33 & 0.013 & 1 \\
11.89 & 0.704 & 1.329 & 0.014 & 1 \\
12.59 & 0.746 & 1.273 & 0.013 & 1 \\
13.34 & 0.79 & 1.211 & 0.012 & 1 \\
14.13 & 0.837 & 1.159 & 0.012 & 1 \\
14.96 & 0.887 & 1.157 & 0.012 & 1.1 \\
15.85 & 0.939 & 1.135 & 0.012 & 1.1 \\
16.79 & 0.995 & 1.146 & 0.012 & 1.1 \\
17.78 & 1.05 & 1.161 & 0.012 & 1.1 \\
18.84 & 1.12 & 1.184 & 0.013 & 1.1 \\
19.95 & 1.18 & 1.17 & 0.013 & 1.1 \\
21.13 & 1.25 & 1.161 & 0.012 & 1.1 \\
22.39 & 1.33 & 1.128 & 0.012 & 1.1 \\
23.71 & 1.41 & 1.117 & 0.012 & 1.1 \\
25.12 & 1.49 & 1.098 & 0.012 & 1.1 \\
26.61 & 1.58 & 1.1 & 0.012 & 1.1 \\
28.18 & 1.67 & 1.076 & 0.012 & 1.1 \\
29.85 & 1.77 & 1.105 & 0.013 & 1.2 \\
31.62 & 1.87 & 1.092 & 0.012 & 1.1 \\
33.50 & 1.98 & 1.091 & 0.012 & 1.1 \\
35.48 & 2.1 & 1.103 & 0.013 & 1.1 \\
37.58 & 2.23 & 1.097 & 0.012 & 1.1 \\
39.81 & 2.36 & 1.105 & 0.013 & 1.2 \\
42.17 & 2.5 & 1.092 & 0.013 & 1.2 \\
44.67 & 2.65 & 1.092 & 0.013 & 1.2 \\
47.32 & 2.8 & 1.075 & 0.012 & 1.2 \\
50.12 & 2.97 & 1.078 & 0.013 & 1.2 \\
53.09 & 3.15 & 1.086 & 0.012 & 1.1 \\
56.23 & 3.33 & 1.113 & 0.013 & 1.1 \\
59.57 & 3.53 & 1.083 & 0.013 & 1.2 \\
63.10 & 3.74 & 1.118 & 0.013 & 1.2 \\
66.83 & 3.96 & 1.08 & 0.013 & 1.2 \\
70.79 & 4.19 & 1.1 & 0.014 & 1.3 \\
74.99 & 4.44 & 1.11 & 0.014 & 1.3 \\
79.43 & 4.71 & 1.115 & 0.015 & 1.3 \\
84.14 & 4.99 & 1.097 & 0.015 & 1.4 \\
89.13 & 5.28 & 1.092 & 0.015 & 1.3 \\
94.41 & 5.59 & 1.081 & 0.015 & 1.4 \\
\end{longtable}

\section{Cross-section Ratio Partial Uncertainties Table}
\label{app:uncertResults}
The fission cross-section ratio partial uncertainties of \pu{} relative to \u{} over a neutron energy range of 0.1 -- 100~MeV are given here. The uncertainties are reported as percentage of the cross-section ratio results in Table~\ref{table:results} of \ref{app:Results}. The bin structure is logarithmic.  The reported energy bin value is the lower edge of the bin.  The bin widths are the same as reported in Table~\ref{table:results}. The partial uncertainties listed are $\sigma_{C_{ff}}$ statistical , $\sigma_v$ from the variational analysis, $\sigma_{C_w}$ is the wraparound correction, $\sigma_{\epsilon_{ff}}$ is the efficiency correction, $\sigma_{C_b}$ is the contamination correction, $\sigma_{\phi_{XY}}$ is the overlap correction and $\sigma_{\kappa}$ is the beam attenuation and scattering correction.
The normalization correction uncertainty is 0.79\% as discussed in Sec.~\ref{sec:normalization}. The normalization uncertainty is energy independent and therefore not listed in the table.

\begin{longtable}{c|c|c|c|c|c|c|c}
\caption{$^{239}$Pu(n,f)/$^{235}$U(n,f) cross-section ratio partial uncertainties (\%). 
}
\label{table:uncertResults}\\
\hline\hline
Energy Bin (MeV) & $\sigma_{C_{ff}}$ & $\sigma_v$ & $\sigma_{C_w}$ & $\sigma_{\epsilon_{ff}}$ & $\sigma_{C_{b}}$ & $\sigma_{\phi_{XY}}$ & $\sigma_{\kappa}$\\
\hline
\endfirsthead
\hline\hline
Energy Bin (MeV) & $\sigma_{C_{ff}}$ & $\sigma_v$ & $\sigma_{C_w}$ & $\sigma_{\epsilon_{ff}}$ & $\sigma_{C_{b}}$ & $\sigma_{\phi_{XY}}$ & $\sigma_{\kappa}$\\
\hline
\endhead
\hline\hline
\endfoot
0.1059 & 1.6 & 1.5 & 0.6 & 0.19 & 0.099 & 0.54 & 0.14 \\
0.1122 & 1.6 & 1.7 & 0.54 & 0.22 & 0.12 & 0.58 & 0.13 \\
0.1189 & 1.4 & 1 & 0.45 & 0.21 & 0.13 & 0.48 & 0.14 \\
0.1259 & 1.4 & 1.3 & 0.41 & 0.23 & 0.11 & 0.46 & 0.12 \\
0.1334 & 1.3 & 1.1 & 0.35 & 0.2 & 0.082 & 0.48 & 0.12 \\
0.1413 & 1.3 & 0.99 & 0.35 & 0.2 & 0.057 & 0.53 & 0.14 \\
0.1496 & 1.1 & 0.83 & 0.28 & 0.2 & 0.057 & 0.47 & 0.12 \\
0.1585 & 1.1 & 1.2 & 0.26 & 0.2 & 0.064 & 0.38 & 0.12 \\
0.1679 & 1.1 & 1 & 0.24 & 0.22 & 0.066 & 0.46 & 0.11 \\
0.1778 & 1 & 0.82 & 0.21 & 0.22 & 0.06 & 0.39 & 0.098 \\
0.1884 & 1 & 0.82 & 0.22 & 0.2 & 0.053 & 0.41 & 0.11 \\
0.1995 & 0.93 & 0.76 & 0.18 & 0.2 & 0.051 & 0.37 & 0.098 \\
0.2113 & 0.91 & 0.74 & 0.17 & 0.2 & 0.053 & 0.38 & 0.083 \\
0.2239 & 0.86 & 0.64 & 0.16 & 0.19 & 0.055 & 0.34 & 0.084 \\
0.2371 & 0.84 & 0.65 & 0.15 & 0.19 & 0.055 & 0.33 & 0.09 \\
0.2512 & 0.81 & 0.57 & 0.14 & 0.2 & 0.053 & 0.32 & 0.097 \\
0.2661 & 0.76 & 0.54 & 0.13 & 0.21 & 0.05 & 0.33 & 0.064 \\
0.2818 & 0.75 & 0.45 & 0.12 & 0.19 & 0.049 & 0.31 & 0.07 \\
0.2985 & 0.69 & 0.45 & 0.1 & 0.21 & 0.049 & 0.34 & 0.074 \\
0.3162 & 0.69 & 0.4 & 0.1 & 0.19 & 0.05 & 0.31 & 0.068 \\
0.3350 & 0.63 & 0.36 & 0.089 & 0.23 & 0.049 & 0.26 & 0.064 \\
0.3548 & 0.63 & 0.34 & 0.082 & 0.22 & 0.048 & 0.28 & 0.07 \\
0.3758 & 0.61 & 0.34 & 0.077 & 0.2 & 0.047 & 0.25 & 0.088 \\
0.3981 & 0.59 & 0.31 & 0.074 & 0.18 & 0.048 & 0.25 & 0.058 \\
0.4217 & 0.58 & 0.32 & 0.073 & 0.19 & 0.048 & 0.25 & 0.062 \\
0.4467 & 0.55 & 0.37 & 0.067 & 0.19 & 0.049 & 0.26 & 0.054 \\
0.4732 & 0.54 & 0.26 & 0.062 & 0.2 & 0.05 & 0.25 & 0.056 \\
0.5012 & 0.5 & 0.26 & 0.057 & 0.2 & 0.05 & 0.23 & 0.056 \\
0.5309 & 0.5 & 0.29 & 0.048 & 0.21 & 0.05 & 0.25 & 0.056 \\
0.5623 & 0.49 & 0.24 & 0.046 & 0.23 & 0.05 & 0.22 & 0.066 \\
0.5957 & 0.47 & 0.21 & 0.04 & 0.2 & 0.05 & 0.2 & 0.052 \\
0.6310 & 0.45 & 0.24 & 0.042 & 0.21 & 0.05 & 0.23 & 0.058 \\
0.6683 & 0.45 & 0.25 & 0.042 & 0.18 & 0.05 & 0.19 & 0.052 \\
0.7079 & 0.44 & 0.22 & 0.037 & 0.19 & 0.049 & 0.2 & 0.054 \\
0.7499 & 0.43 & 0.2 & 0.037 & 0.21 & 0.049 & 0.21 & 0.046 \\
0.7943 & 0.42 & 0.2 & 0.033 & 0.22 & 0.05 & 0.2 & 0.052 \\
0.8414 & 0.41 & 0.22 & 0.03 & 0.2 & 0.049 & 0.22 & 0.039 \\
0.8913 & 0.41 & 0.24 & 0.027 & 0.21 & 0.048 & 0.18 & 0.038 \\
0.9441 & 0.4 & 0.22 & 0.026 & 0.21 & 0.046 & 0.2 & 0.04 \\
1.000 & 0.4 & 0.22 & 0.024 & 0.22 & 0.046 & 0.2 & 0.039 \\
1.059 & 0.39 & 0.24 & 0.023 & 0.2 & 0.046 & 0.19 & 0.04 \\
1.122 & 0.39 & 0.21 & 0.021 & 0.2 & 0.046 & 0.18 & 0.046 \\
1.189 & 0.39 & 0.21 & 0.02 & 0.19 & 0.045 & 0.19 & 0.041 \\
1.259 & 0.38 & 0.26 & 0.022 & 0.18 & 0.044 & 0.19 & 0.042 \\
1.334 & 0.38 & 0.22 & 0.021 & 0.21 & 0.043 & 0.23 & 0.034 \\
1.413 & 0.38 & 0.22 & 0.018 & 0.2 & 0.042 & 0.2 & 0.042 \\
1.496 & 0.38 & 0.21 & 0.017 & 0.19 & 0.041 & 0.17 & 0.082 \\
1.585 & 0.38 & 0.19 & 0.018 & 0.22 & 0.04 & 0.19 & 0.039 \\
1.679 & 0.38 & 0.28 & 0.017 & 0.23 & 0.039 & 0.18 & 0.04 \\
1.778 & 0.38 & 0.19 & 0.017 & 0.17 & 0.038 & 0.18 & 0.047 \\
1.884 & 0.38 & 0.23 & 0.017 & 0.19 & 0.036 & 0.18 & 0.036 \\
1.995 & 0.37 & 0.23 & 0.015 & 0.21 & 0.035 & 0.17 & 0.041 \\
2.113 & 0.38 & 0.21 & 0.014 & 0.18 & 0.034 & 0.18 & 0.046 \\
2.239 & 0.38 & 0.22 & 0.013 & 0.21 & 0.033 & 0.19 & 0.037 \\
2.371 & 0.39 & 0.2 & 0.015 & 0.21 & 0.032 & 0.19 & 0.041 \\
2.512 & 0.39 & 0.21 & 0.014 & 0.2 & 0.031 & 0.19 & 0.032 \\
2.661 & 0.39 & 0.2 & 0.013 & 0.2 & 0.029 & 0.22 & 0.039 \\
2.818 & 0.4 & 0.21 & 0.014 & 0.17 & 0.028 & 0.2 & 0.048 \\
2.985 & 0.4 & 0.24 & 0.013 & 0.2 & 0.027 & 0.17 & 0.04 \\
3.162 & 0.4 & 0.22 & 0.013 & 0.21 & 0.026 & 0.2 & 0.04 \\
3.350 & 0.41 & 0.22 & 0.014 & 0.18 & 0.025 & 0.22 & 0.038 \\
3.548 & 0.42 & 0.17 & 0.014 & 0.2 & 0.023 & 0.22 & 0.046 \\
3.758 & 0.41 & 0.23 & 0.013 & 0.19 & 0.022 & 0.21 & 0.042 \\
3.981 & 0.43 & 0.23 & 0.012 & 0.19 & 0.021 & 0.21 & 0.042 \\
4.217 & 0.44 & 0.2 & 0.012 & 0.2 & 0.02 & 0.16 & 0.04 \\
4.467 & 0.45 & 0.24 & 0.013 & 0.2 & 0.02 & 0.22 & 0.037 \\
4.732 & 0.46 & 0.18 & 0.013 & 0.18 & 0.02 & 0.22 & 0.036 \\
5.012 & 0.47 & 0.19 & 0.014 & 0.19 & 0.021 & 0.2 & 0.04 \\
5.309 & 0.48 & 0.21 & 0.013 & 0.18 & 0.022 & 0.21 & 0.036 \\
5.623 & 0.49 & 0.23 & 0.013 & 0.16 & 0.023 & 0.22 & 0.037 \\
5.957 & 0.48 & 0.2 & 0.012 & 0.19 & 0.022 & 0.22 & 0.041 \\
6.310 & 0.46 & 0.21 & 0.011 & 0.2 & 0.02 & 0.24 & 0.039 \\
6.683 & 0.45 & 0.21 & 0.0099 & 0.23 & 0.018 & 0.19 & 0.045 \\
7.079 & 0.44 & 0.21 & 0.0091 & 0.2 & 0.017 & 0.24 & 0.054 \\
7.499 & 0.45 & 0.2 & 0.009 & 0.2 & 0.016 & 0.21 & 0.05 \\
7.943 & 0.46 & 0.21 & 0.0091 & 0.22 & 0.015 & 0.24 & 0.057 \\
8.414 & 0.47 & 0.2 & 0.0092 & 0.2 & 0.015 & 0.22 & 0.052 \\
8.913 & 0.48 & 0.22 & 0.0096 & 0.22 & 0.014 & 0.22 & 0.052 \\
9.441 & 0.5 & 0.22 & 0.0099 & 0.19 & 0.015 & 0.23 & 0.049 \\
10.00 & 0.52 & 0.22 & 0.01 & 0.21 & 0.016 & 0.21 & 0.05 \\
10.59 & 0.55 & 0.2 & 0.011 & 0.18 & 0.018 & 0.27 & 0.052 \\
11.22 & 0.57 & 0.23 & 0.012 & 0.21 & 0.022 & 0.28 & 0.054 \\
11.89 & 0.59 & 0.25 & 0.012 & 0.23 & 0.026 & 0.29 & 0.068 \\
12.59 & 0.59 & 0.25 & 0.012 & 0.18 & 0.027 & 0.3 & 0.064 \\
13.34 & 0.6 & 0.19 & 0.012 & 0.2 & 0.029 & 0.29 & 0.061 \\
14.13 & 0.61 & 0.23 & 0.012 & 0.21 & 0.031 & 0.31 & 0.067 \\
14.96 & 0.61 & 0.21 & 0.012 & 0.22 & 0.031 & 0.34 & 0.061 \\
15.85 & 0.63 & 0.25 & 0.012 & 0.2 & 0.031 & 0.3 & 0.067 \\
16.79 & 0.65 & 0.21 & 0.012 & 0.2 & 0.029 & 0.34 & 0.062 \\
17.78 & 0.64 & 0.24 & 0.012 & 0.22 & 0.023 & 0.3 & 0.074 \\
18.84 & 0.66 & 0.22 & 0.012 & 0.18 & 0.016 & 0.33 & 0.086 \\
19.95 & 0.68 & 0.22 & 0.012 & 0.2 & 0.016 & 0.29 & 0.078 \\
21.13 & 0.68 & 0.25 & 0.012 & 0.21 & 0.016 & 0.34 & 0.075 \\
22.39 & 0.68 & 0.19 & 0.012 & 0.17 & 0.016 & 0.33 & 0.081 \\
23.71 & 0.68 & 0.21 & 0.011 & 0.2 & 0.016 & 0.36 & 0.074 \\
25.12 & 0.69 & 0.21 & 0.011 & 0.17 & 0.014 & 0.38 & 0.079 \\
26.61 & 0.69 & 0.2 & 0.011 & 0.23 & 0.011 & 0.36 & 0.074 \\
28.18 & 0.7 & 0.19 & 0.011 & 0.2 & 0.007 & 0.36 & 0.078 \\
29.85 & 0.7 & 0.21 & 0.011 & 0.2 & 0.0056 & 0.36 & 0.093 \\
31.62 & 0.7 & 0.22 & 0.011 & 0.21 & 0.0056 & 0.34 & 0.087 \\
33.50 & 0.69 & 0.19 & 0.0098 & 0.2 & 0.0056 & 0.3 & 0.087 \\
35.48 & 0.7 & 0.3 & 0.01 & 0.22 & 0.0056 & 0.34 & 0.081 \\
37.58 & 0.71 & 0.26 & 0.0096 & 0.18 & 0.0056 & 0.34 & 0.089 \\
39.81 & 0.72 & 0.2 & 0.0097 & 0.19 & 0.0056 & 0.35 & 0.09 \\
42.17 & 0.73 & 0.2 & 0.0095 & 0.23 & 0.0056 & 0.38 & 0.08 \\
44.67 & 0.73 & 0.26 & 0.0095 & 0.2 & 0.0056 & 0.31 & 0.085 \\
47.32 & 0.75 & 0.22 & 0.0094 & 0.2 & 0.0056 & 0.35 & 0.11 \\
50.12 & 0.77 & 0.26 & 0.0095 & 0.18 & 0.0056 & 0.35 & 0.1 \\
53.09 & 0.76 & 0.2 & 0.0094 & 0.17 & 0.0056 & 0.39 & 0.11 \\
56.23 & 0.78 & 0.21 & 0.0092 & 0.21 & 0.0056 & 0.41 & 0.088 \\
59.57 & 0.79 & 0.2 & 0.0093 & 0.18 & 0.0056 & 0.39 & 0.095 \\
63.10 & 0.8 & 0.24 & 0.0092 & 0.22 & 0.0056 & 0.38 & 0.11 \\
66.83 & 0.81 & 0.24 & 0.0095 & 0.2 & 0.0056 & 0.39 & 0.094 \\
70.79 & 0.85 & 0.23 & 0.0096 & 0.21 & 0.0056 & 0.37 & 0.1 \\
74.99 & 0.87 & 0.27 & 0.0098 & 0.2 & 0.0056 & 0.49 & 0.097 \\
79.43 & 0.9 & 0.2 & 0.0097 & 0.17 & 0.0056 & 0.42 & 0.094 \\
84.14 & 0.91 & 0.38 & 0.01 & 0.2 & 0.0056 & 0.42 & 0.098 \\
89.13 & 0.94 & 0.24 & 0.011 & 0.22 & 0.0056 & 0.51 & 0.095 \\
94.41 & 0.99 & 0.27 & 0.011 & 0.18 & 0.0056 & 0.45 & 0.1 \\
\end{longtable}

\end{document}